\RequirePackage{lineno}
\documentclass[aapm,mph,unsortedaddress,citeautoscript,showkeys]{revtex4} 
\usepackage[english]{babel}
\usepackage{amsmath,amsfonts,amssymb}
\usepackage{graphicx}
\usepackage{setspace}
\usepackage{tocloft}
\usepackage{siunitx}
\usepackage{bm}
\usepackage{pgfplots}
\pgfplotsset{compat=newest}
\usepackage{tikz}
\usetikzlibrary{positioning,pgfplots.groupplots,matrix}
\usetikzlibrary{fit}
\usepackage{tabularx}
\usepackage{setspace}
\setcitestyle{super}

\let\oldequation\equation
\let\oldendequation\endequation
\renewenvironment{equation}
  {\linenomathNonumbers\oldequation}
  {\oldendequation\endlinenomath} 

\def\i{\mbox{\small{\rm i}}}

\newcommand{\zb}[1]{\mbox{\boldmath{${#1}$}}}

\newcommand{\adj}{{\ensuremath{\mathsf{H}}}}

\begin{document} 
%\linenumbers
\setstretch{1.0}
\nocite{apsrev41Control}
\title{Bimodal intravascular volumetric imaging combining OCT and MPI}

\author{Sarah Latus$^\text{a,*}$, Florian Griese$^\text{b,c,*,+}$, Matthias Schl\"uter$^\text{a}$, Christoph Otte$^\text{a}$, Martin M\"oddel$^\text{b,c}$, Matthias Graeser$^\text{b,c}$, Thore Saathoff$^\text{a}$, Tobias Knopp$^\text{b,c}$, Alexander Schlaefer$^\text{a}$}
\address{$^\text{a}$ Institute of Medical Technology, Hamburg University of Technology, Hamburg, 21073, Germany}
\address{$^\text{b}$Section for Biomedical Imaging, University Medical Center Hamburg-Eppendorf, Hamburg, 20246, Germany}
\address{$^\text{c}$Institute for Biomedical Imaging, Hamburg University of Technology, Hamburg, 21073, Germany}
\address{$^\text{*}$Authors contributed equally}
\address{$^\text{+}$Corresponding author: Florian Griese, f.griese@uke.de}
%\affiliation{$^\text{a}$Institute of Medical Technology, Hamburg University of Technology, Hamburg, 21073, Germany}
%\affiliation{$^\text{b}$Section for Biomedical Imaging, University Medical Center Hamburg-Eppendorf, Hamburg, 20246, Germany}
%\affiliation{$^\text{c}$Institute for Biomedical Imaging, Hamburg University of Technology, Hamburg, 21073, Germany}
%\affiliation{$^\text{*}$Authors contributed equally}
%\affiliation{$^\text{+}$Corresponding author: Florian Griese, f.griese@uke.de}

\begin{abstract}
%\linenumbers
\noindent \textbf{Purpose:}
Intravascular optical coherence tomography (IVOCT) is a catheter based image modality allowing for high resolution imaging of vessels. It is based on a fast sequential acquisition of A-scans with an axial spatial resolution in the range of \SI{5} to \SI{10}{\micro\meter}, i.e., one order of magnitude higher than in conventional methods like intravascular ultrasound or computed tomography angiography.
However, position and orientation of the catheter in patient coordinates cannot be obtained from the IVOCT measurements alone. Hence, the pose of the catheter needs to be established to correctly reconstruct the three-dimensional vessel shape. Magnetic particle imaging (MPI) is a three-dimensional tomographic, tracer-based and radiation-free image modality providing high temporal resolution with unlimited penetration depth. Volumetric MPI images are angiographic and hence suitable to complement IVOCT as a co-modality. We study simultaneous bimodal IVOCT MPI imaging with the goal of estimating the IVOCT pullback path based on the 3D MPI data.

\noindent\textbf{Methods:}
We present a setup to study and evaluate simultaneous IVOCT and MPI image acquisition of differently shaped vessel phantoms. First, the influence of the MPI tracer concentration on the optical properties required for IVOCT is analyzed. Second, using a concentration allowing for simultaneous imaging, IVOCT and MPI image data is acquired sequentially and simultaneously. Third,
the luminal centerline is established from the MPI image volumes and used to estimate the catheter pullback trajectory for IVOCT image reconstruction. The image volumes are compared to the known shape of the phantoms.

\noindent\textbf{Results:}
We were able to identify a suitable MPI tracer concentration of $\SI{2.5}{\milli\mol\per\liter}$ with negligible influence on the IVOCT signal. The pullback trajectory estimated from MPI agrees well with the centerline of the phantoms. Its mean absolute error ranges from $\SI{0.27}{\milli\meter}$ to $\SI{0.28}{\milli\meter}$ and from $\SI{0.25}{\milli\meter}$ to $\SI{0.28}{\milli\meter}$ for sequential and simultaneous measurements, respectively. Likewise, reconstructing the shape of the vessel phantoms works well with mean absolute errors for the diameter 
ranging from $\SI{0.11}{\milli\meter}$ to $\SI{0.21}{\milli\meter}$ and from $\SI{0.06}{\milli\meter}$ to $\SI{0.14}{\milli\meter}$ for sequential and simultaneous measurements, respectively.

\noindent\textbf{Conclusions:} 
MPI can be used in combination with IVOCT to estimate the catheter trajectory and the vessel shape with high precision and without ionizing radiation. 
\end{abstract}

% Include a list of up to six keywords after the abstract

\maketitle

\begin{spacing}{1.0}   % use double spacing for rest of manuscript

\section{Introduction}\label{sect:intro}

Optical coherence tomography (OCT) is a tomographic imaging method providing very high spatial resolution between \SI{5}{} and \SI{10}{\micro\meter} \cite{huang1991optical}. OCT has a high temporal resolution of above \SI{1}{\mega\hertz} \cite{Wang2015,pfeiffer2018} for its 1D imaging depth profiles called A-scan. Since OCT has only a limited penetration depth in tissue in the range of \SI{1}{} to \SI{2}{\milli\meter}, its main application is near surface imaging such as in ophthalmology, dermatology, odontology, and intravascular imaging.
In order to use OCT for intravascular imaging, the OCT fiber is integrated into a catheter and scanning is realized by rotating a prism at the fiber tip to acquire A-scans virtually orthogonal to the catheter. Hence, intravascular OCT (IVOCT) allows imaging vessel boundaries in cardiovascular applications and offers a spatial resolution that is more than one magnitude higher than intravascular ultrasound (IVUS), computed tomography angiography (CTA), or magnetic resonance angiography (MRA). Primarily, IVOCT is used to assess the vascular wall morphology, e.g., for diagnosis of atherosclerosis and thin-cap fibroatheroma \cite{tearney2012}.

While IVOCT offers very high spatio-temporal resolution, one major challenge is that the position and orientation of the A-scans in the global patient coordinate frame is unknown. A simple model assumes motion along a line and uses the known rotation and the pullback speed to calculate the orientation and position of the A-scans. This, however, neglects that the vessel is not perfectly straight and it does also not take non-linear movements during the pullback and non-uniform rotations into account. In order to estimate the path of the catheter movement -- the so-called pullback trajectory -- IVOCT can be combined with a second image modality showing the catheter and vessel in patient coordinates. Digital subtraction angiography (DSA) is often used for this purpose \cite{DeCock2014}. 
But DSA only provides 2D projection images and in order to get a 3D trajectory two DSA images have to be taken from different angles. These 2D trajectories from both DSA images can then be extruded perpendicularly to the angiographic planes forming two surfaces. From the intersection of the two surfaces the 3D trajectory can be estimated \cite{bourantas2013new}. In principle, this can be achieved from successive monoscopic images. Additionally, aspects like bi-planar acquisition and compensation for cardiac and respiratory should be considered in that case\cite{lessmann2014}. Other drawbacks of DSA are its use of ionizing radiation and a contrast agent (iodine) that cannot be used in all patients. e.g., in the case of kidney diseases \cite{katzberg2006contrast,mccullough1997acute,mccullough2008contrast}. 

The purpose of this work is to investigate an alternative to DSA with the new tracer-based imaging method magnetic particle imaging (MPI), since MPI is predestined for vascular imaging with its 3D information, high temporal resolution and unlimited penetration depth \cite{knopp2017recent, knopp2017magnetic}. MPI does not use radiation but only applies static and oscillating magnetic fields that are safe for the patient, if the oscillating fields strength is within the safety constraints of the peripheral nerve stimulation (PNS) \cite{saritas_magnetostimulation_2013} and the limits for the specific absorption rate (SAR) are satisfied \cite{Bohnert2008,Bohnert2009,schmale2015mpi}. Super-paramagnetic iron-oxide particles (SPIOs) provide the contrast in MPI, which have been already used extensively for the human body in MRI \cite{neuwelt2009ultrasmall,lu2010fda}. MPI has proven to be suited for vascular imaging \cite{haegele2012magnetic,haegele2013toward,salamon2016magnetic} while offering a spatial resolution in the range of \SI{0.5}{} to \SI{5}{\milli\meter}. This suffices to quantify vascular stenosis \cite{10.1371/journal.pone.0168902,storath2017edge}. Catheters and guide wires can be coated with magnetic nanoparticles enabling their visualization in MPI \cite{haegele2016magnetic}. Using multi-spectral reconstruction techniques \cite{haegele2016multi}, it is even possible to simultaneously image the coated instrument/catheter and the blood pool tracer making navigation tasks possible. Using long circulating tracers \cite{kaul2017vitro}, the vessel tree can be visualized over several hours which is a huge advantage over the use of tracers in the DSA setting where iodine has to be regularly injected. Since MPI imaging can be performed in 3D, MPI is advantageous for determining the pose of vessels in 3D space. 

To establish the feasibility of combining OCT and MPI, we first analyze the influence of MPI tracers on the OCT signal to identify a suitable concentration allowing for bimodal imaging. Second, we describe an approach to reconstruct volumetric IVOCT images based on the MPI data. Third, we present results for an experimental evaluation of the proposed method. For comparison, we consider different assumptions regarding the vessel shape when reconstructing image volumes for vessel phantoms of different shape. Our results indicate that improved volumetric imaging is feasible for both sequential and simultaneous IVOCT and MPI imaging.

\section{Material and Methods}

\subsection{Optimization of the imaging protocol}

\begin{figure}[htp]
\centering
\includegraphics[width=1.0\linewidth]{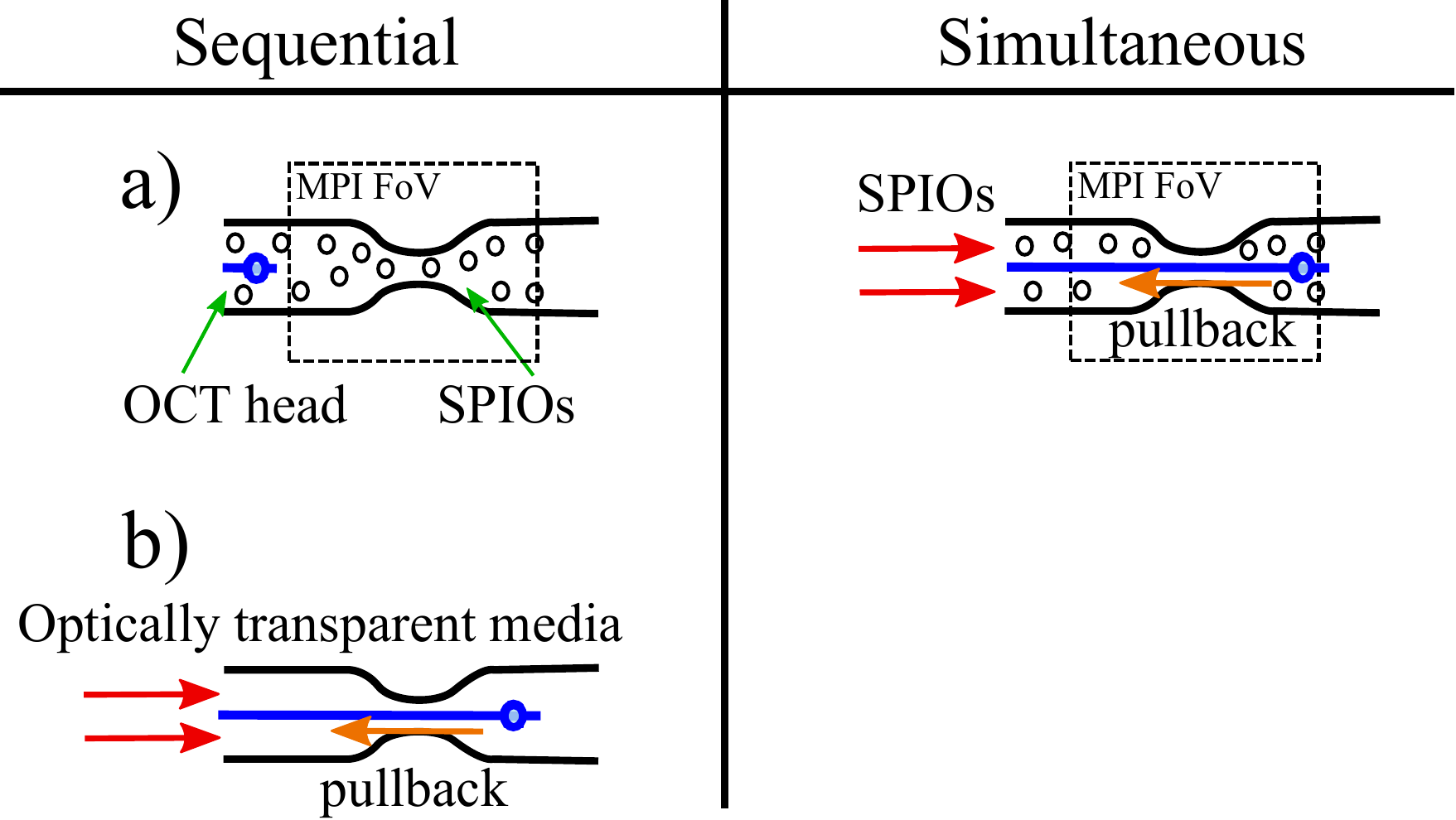}
\caption{Scenarios for bimodal IVOCT MPI imaging. In the sequential scenario  the MPI dataset a) is acquired prior to the IVOCT data b). In the simultaneous scenario both MPI and IVOCT measurements are performed simultaneously.}
\label{fig:Figure01}
\end{figure}

Generally, we consider the situation where the IVOCT catheter is already placed at the position in the vessel tree where a high resolution OCT measurement is planned. Placement of the catheter could be realized by labeling the catheter and the guidewire with an MPI visible marker and performing an interventional navigation procedure as outlined in \cite{salamon2016magnetic}. In principle, the imaging could be realized sequentially, e.g., by first studying the vessel shape using MPI and subsequently acquiring the IVOCT A-scans. However, with a suitable tracer concentration it would be possible to acquire the images simultaneously. This would be preferable to monitor the actual catheter motion during pullback. The two approaches are sketched in Fig.~\ref{fig:Figure01}.

In the sequential imaging mode, first the MPI measurement is performed while the IVOCT catheter is not pulled back. When using a long circulating MPI tracer, no additional tracer has to be injected as the vessel is already visible in MPI. After the MPI image was taken, the IVOCT catheter is fed forward and the blood has to be flushed with an optically transparent solution because otherwise the OCT beam would be absorbed by the blood and would not reach the vessel boundary. While in clinical practice iodine is used for flushing the vessel \cite{tearney2012} in order to simultaneously measure it with DSA, in the bimodal IVOCT MPI setting one could use a NaCl solution. During flushing, the IVOCT catheter is pulled back and the OCT measurement is performed.

In the simultaneous imaging mode, the MPI measurement is performed during the pullback. In general, MPI is based on magnetic particles and the concentration of SPIO has a direct impact on the signal strength. In addition to that, SPIO particles act as scatterers and the tracer therefore changes the attenuation of the infrared light used for IVOCT. Hence, higher SPIO concentrations increase the MPI signal and decrease the OCT signal, and vice versa. To establish a suitable trade-off allowing for good signal-to-noise ratios (SNR) for both modalities we study different SPIO concentrations.

\subsection{Experimental Setup}

The bimodal imaging experiments are performed with a pre-clinical MPI scanner \cite{BrukerScanner} (MPI PreClinical, Bruker). The scanner has a bore diameter of \SI{118}{\milli\meter} and can be used for imaging small animals such as rats and mice. For the IVOCT measurements a custom made setup connecting a catheter (Dragonfly Duo Kit, Abbott) with an outer diameter of \SI{0.9}{\milli\meter} to a spectral-domain OCT device (Telesto I, Thorlabs) is used. The setup allows direct control of pullback and rotation and access to timestamps and raw OCT spectra.

\begin{figure}
\centering
\includegraphics[width=1.0\linewidth]{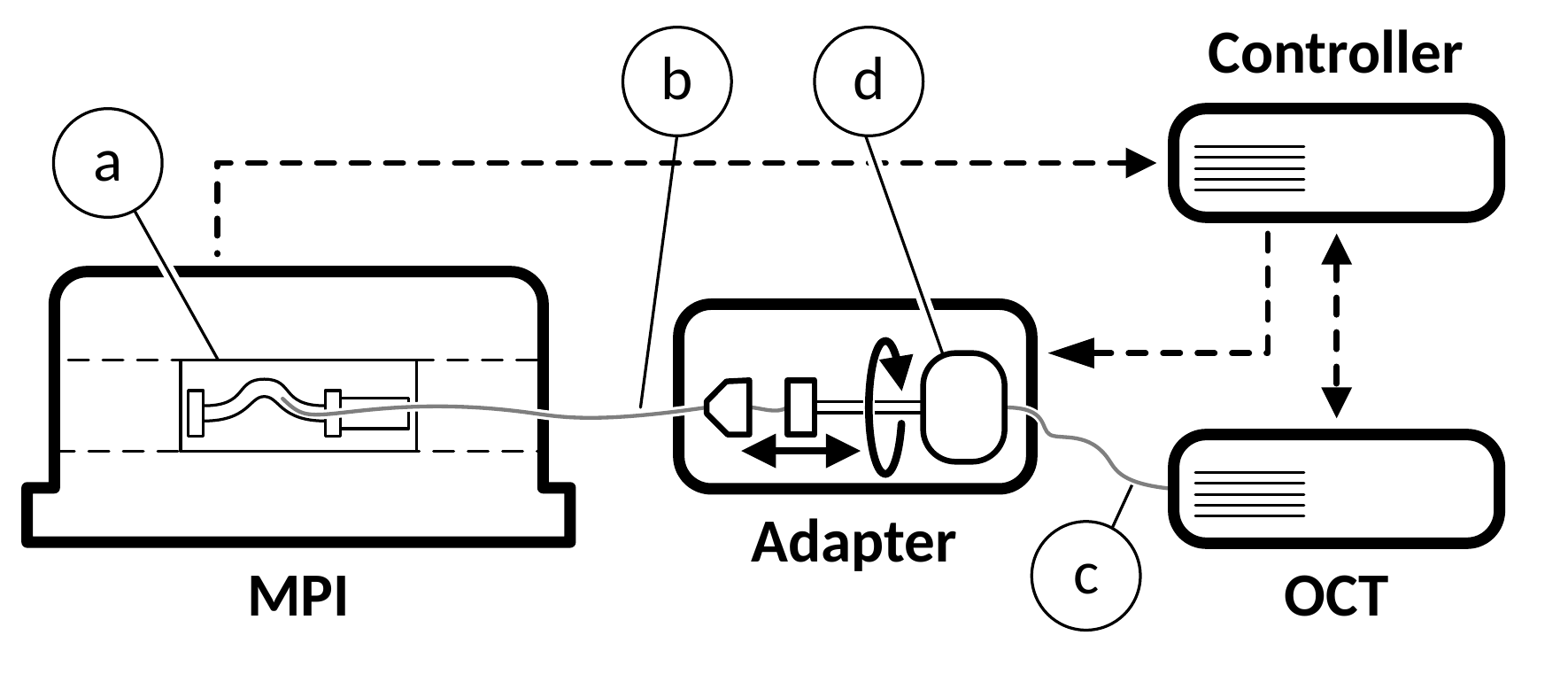}
\caption{The overall setup including MPI, catheter adapter, OCT, and a controller. The phantom (a) is placed in the MPI system and the catheter extending into the phantom (b) is connected to the adapter. The adapter realizes rotation and translation of the fiber inside the catheter. It is connected to the OCT by a second fiber (c) and a rotary coupler (d). Dashed lines illustrate communication between the components.}
\label{fig:Figure02}
\end{figure}

\begin{figure}[thb]
\begin{tabular}{cc}		
\includegraphics[width=0.48\linewidth]{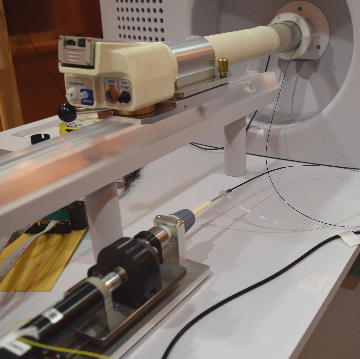} &
\includegraphics[width=0.48\linewidth]{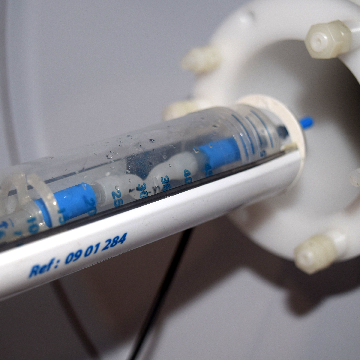}
\end{tabular}
\caption{Left: MPI sample holder and the custom made adapter in front of the MPI system. Right: A vessel phantom with an IVOCT catheter inside before being inserted into the MPI system.}
\label{fig:Figure03}
\end{figure}

The combination of both imaging devices is sketched in Fig.~\ref{fig:Figure02}. Phantoms are placed in the MPI device and the catheter is fed into the phantom. A custom made adapter connects the fiber in the catheter with a fiber attached to the OCT device. The two fibers are connected through a rotary coupler and the adapter uses a DC motor to rotate and a stepper motor to pull back the fiber inside the catheter. The control software of the IVOCT measurement setup, which is implemented in C++ and running on a dedicated controller PC, reads a trigger when the MPI is starting its image sequence and in turn controls the catheter motion and the OCT image acquisition. The MPI device with the sample and the custom made adapter are also shown in Fig.~\ref{fig:Figure03}.

\subsubsection{Vessel Phantoms}
\begin{figure}[!tb]
\begin{tikzpicture}
\node at (-6,0) {\includegraphics[width=0.29\linewidth]{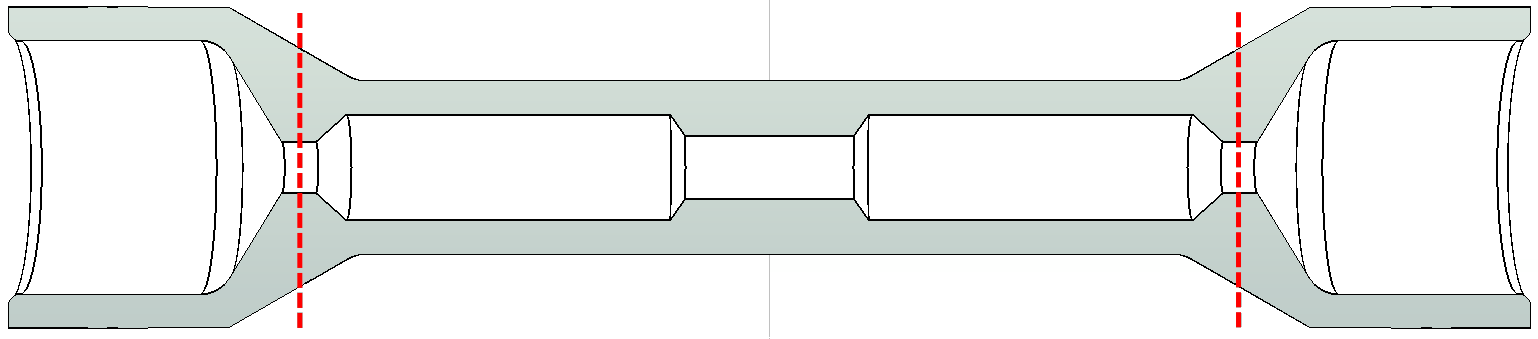}};
\node at (0,0) {\includegraphics[width=0.29\linewidth]{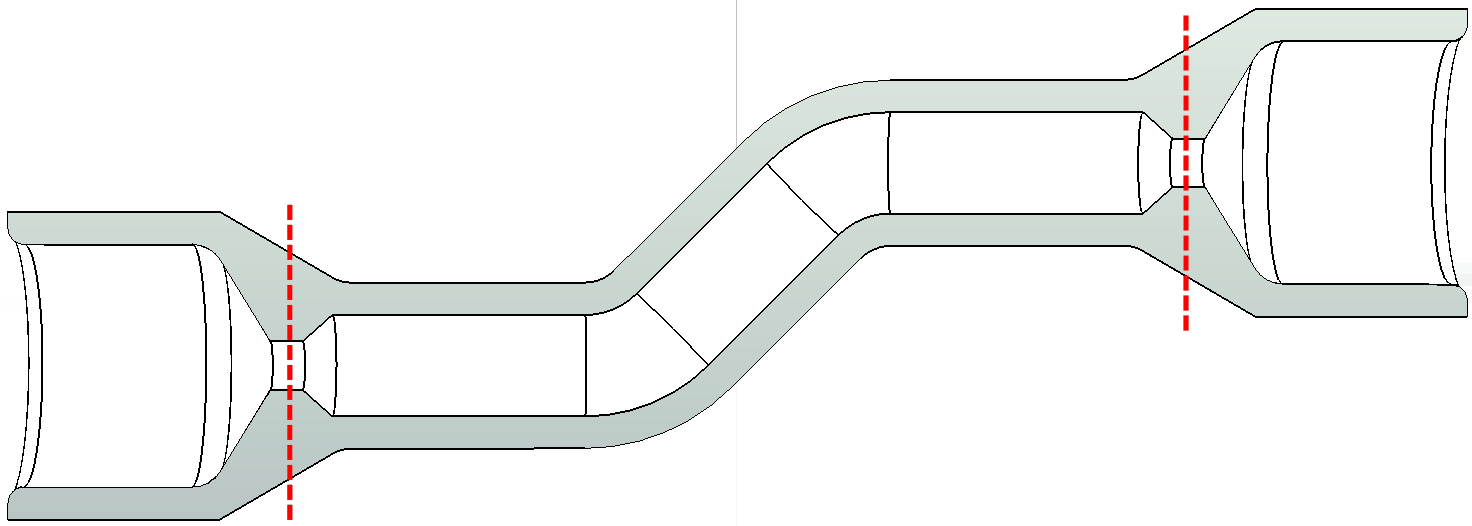}};
\node at (6,0) {\includegraphics[width=0.29\linewidth]{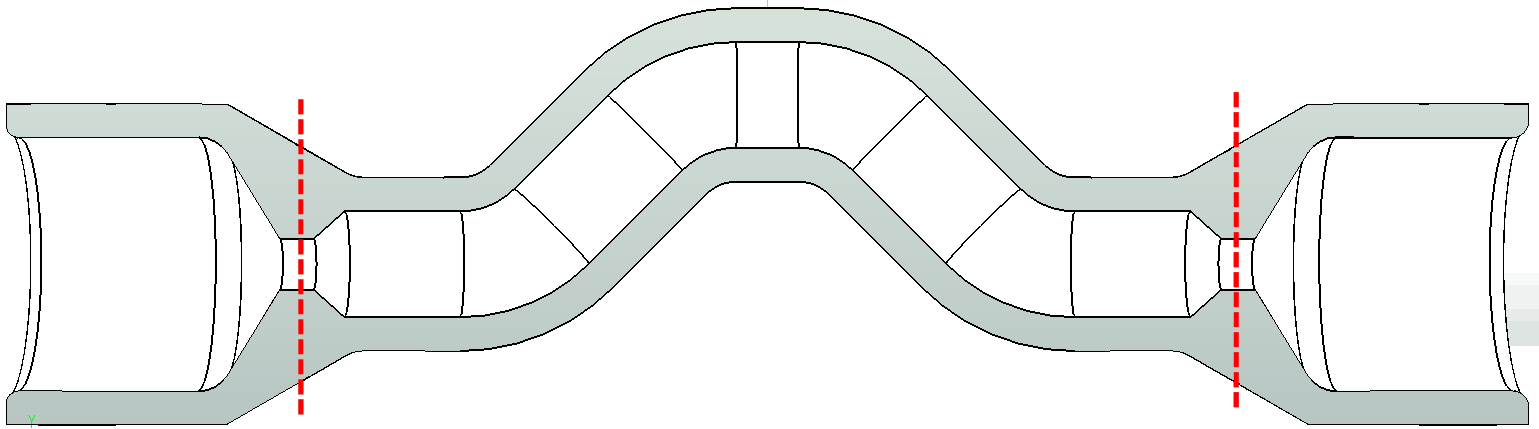}};

\node at (-6,1.5) {Stenosis};
\node at (0,1.5) {Z-shape};
\node at (6,1.5) {U-shape};

\draw[|-|] (-7.6,-1.25) -- (-4.4,-1.25) node[midway,below]{\SI{25}{\milli\meter}};
\draw[|-|] (-1.6,-1.25) -- (1.6,-1.25) node[midway,below]{\SI{25}{\milli\meter}};
\draw[|-|] (7.6,-1.25) -- (4.4,-1.25) node[midway,below]{\SI{25}{\milli\meter}};
\end{tikzpicture}
\caption{CAD models of the 3D-printed vessel phantoms. The dashed lines with distance $l$=\SI{25}{\milli\meter} indicate the positions of phantom landmarks for later co-registration methods.}
\label{fig:Figure04}
\end{figure}

We have designed three different vessel phantoms shown in Fig.~\ref{fig:Figure04} to mimic relevant vessel shapes: a stenosis phantom and two phantoms with Z-shape and U-shape. We choose a total phantom length of \SI{25}{\milli\meter} and an inner diameter of \SI{2.5}{\milli\meter} (stenosis narrow part \SI{1.5}{\milli\meter}). The landmarks on each side of the phantom (dashed red lines) in Fig.~\ref{fig:Figure04} are used to co-register MPI and IVOCT images and crop the resulting 3D volumes.

All phantoms are 3D printed with a 3D printer based on stereolithography (Form 1+, Formlabs). This printer offers a high resolution of \SI{0.05}{\milli\meter} in all axis.
A white colored resin (FLGPWH01, Formlabs) is used for printing, resulting in solid phantoms with \SI{65}{\mega\pascal} ultimate tensile strength and a Young's Modulus of \SI{2.8}{\giga\pascal} after curing.

\def\tWFactor{0.3}

\subsubsection{MPI Acquisition Parameters}

The preclinical MPI scanner features three orthogonal sinusoidal excitation fields with frequencies of $f_x = \SI{2.5/102}{\mega\hertz}$, $f_y=\SI{2.5/96}{\mega\hertz}$, and $f_z=\SI{2.5/99}{\mega\hertz}$. The excitation amplitude is set to \SI{12}{\milli\tesla} in all three directions. The resulting imaging period length is \SI{21.54}{\milli\second} resulting in a frame rate of \SI{46.43}{\hertz}. The gradient of the selection field responsible for spatial encoding is set to \SI{2.0}{\tesla\per\meter} in $z$-direction and \SI{1.0}{\tesla\per\meter} in the $x$- and $y$-directions. The resulting field of view (FoV) is \SI{24x24x12}{\milli\meter}. For signal reception a custom build receive coil is used \cite{Graeser2017}. The MPI data acquisition is performed with the software Paravision from Bruker.

Prior to image reconstruction, it is necessary to perform a calibration scan, which consists of moving a small delta sample filled with SPIOs through the FoV while measuring the system response at all attended positions. The resulting data forms the MPI system matrix that describes the relation between the induced voltage signal and the particle distribution. In this work, the system matrix is measured at \SI{35x25x13}{} positions covering a volume of \SI{35x25x13}{\milli\meter}. The volume is chosen slightly larger to circumvent artifacts at the FoV boundaries \cite{weber2015artifact}. The delta sample has a size of \SI{1x1x1}{\milli\meter} and is filled with \SI{1}{\micro\liter} Perimag (micromod, Rostock) with an iron concentration of \SI{10}{\milli\mol\per\liter}. The same batch of SPIOs is used in the phantom measurements.

\subsubsection{IVOCT Acquisition Parameters}\label{sec:IVOCTAcq}

The spectral-domain OCT system uses a central wavelength of \SI{1315}{\nano\meter} and provides an A-scan rate of $f_{\text{OCT}} = \SI{91}{\kilo\hertz}$. The axial FoV in air is approximately $d_\text{FoV}=\SI{2.66}{\milli\meter}$ and each A-scan contains $P=512$~pixels.  Assuming a constant refractive index of water $n_{\text{H}_2\text{O}}=\SI{1.33}{}$ yields a  pixel spacing $\Delta = d_\text{FoV}/P/n_{\text{H}_2\text{O}} = \SI{4.0}{\micro\meter}$ for the area between the catheter and the phantom wall.

The custom made catheter adapter is operated at a frequency of approximately $f_\text{rot}=\SI{16.6}{\hertz}$ and a pullback velocity of $v_\text{pull}=\SI{0.75}{\milli\meter\per\second}$. Within the MPI drive-field FoV, the catheter is pulled back over a distance of $s=\SI{25}{\milli\meter}$. This results in an IVOCT measurement time of $t_{\text{meas}}=\SI{33.3}{\second}$ and an acquisition of $M\approx\SI{3}\cdot 10^6{}$~A-scans based on $t_{\text{meas}}=s / v_\text{pull}$ and $M=f_{\text{OCT}} \cdot t_{\text{meas}}$.

\subsection{Data Processing} \label{Sec:DataProcessing}

The MPI and IVOCT data is processed in the following way. First, the MPI and OCT data are reconstructed yielding a volumetric 3D image and a set of A-scans respectively. Second, the A-scans are used to calculate cross-sectional IVOCT images. For assembly of these IVOCT images to a complete volume scan, we compare two state-of-the-art methods and our approach which uses information derived from the MPI volume. 

\subsubsection{MPI Image Reconstruction} 

For MPI image reconstruction a first-order Tikhonov-regularized least-squares approach 
\begin{align}
\underset{\bm c}{\text{argmin}} \quad \Vert \zb S \zb c - \zb u \Vert_2^2 + \lambda \Vert \zb c \Vert_2^2 
\end{align}
is considered, where $\zb S \in \mathbb{C}^{M\times N}$ is the MPI system matrix, $\zb u \in \mathbb{C}^{M}$ is the measurement vector and $\zb c \in \mathbb{C}^{N}$ is the particle-concentration vector. The least-squares problem is iteratively solved using the Kaczmarz method, which is known to converge rapidly for MPI system matrices \cite{Knopp2010e,knopp2016online}. The respective MPI data processing is implemented in our own framework written in Julia.

The reconstruction parameters are optimized based on visual inspection of the reconstructed images. The number of Kaczmarz iteration is set to 1 while the regularization parameter $\lambda$ is set to $\lambda = \lambda_0 \cdot 10^{-3}$, where $\lambda_0 = \text{trace}(\zb S^\adj \zb S)N^{-1}$.

\subsubsection{OCT Image Reconstruction} 

The OCT system is based on a Michelson interferometer. Within the coherence length of the light source, the reflected light from the reference arm and the sample arm interfere and a spectrometer records the resulting spectrum. Given the spectrum of the light source $S(k)$ and a finite number $M$ of ideal reflections in the sample arm, the acquired signal is proportional to
\begin{equation}\label{e:oct}
\begin{split}
I_D(k) = & S(k) \left( R_r + \sum_{m=1}^M R_m \right) \\
&+S(k) \sum_{m=1}^M 2\sqrt{R_r R_m} \left( \cos[2k(z_r - z_m)] \right)\\
&+S(k) \sum_{m=1}^M \sum_{n=1,n\neq m}^M \sqrt{R_m R_n} \left( \cos[2k(z_m - z_n)] \right)
\end{split}
\end{equation}
where $R_i$ is the reflectivity, $z_i$ the path length till reflection, and the index $r$ refers to the reference arm \cite{drexlerfujimoto}. While the first term in Eq.~\eqref{e:oct} mainly represents a static offset related to the used light source, the desired information about interference between reference and sample arm is encoded in the second term. The third term describes the effect that also the different reflections within the sample arm interfere pairwise.

OCT data reconstruction described below is implemented in MATLAB and performed individually for each raw data spectrum measured with the OCT device. First, the offset term in Eq.~\eqref{e:oct} is compensated by background subtraction. As the data is not sampled equidistantly w.r.t.\ wave number $k$, a re-gridding step is then necessary. Finally, apodization using a Hann window and fast Fourier transform are applied to obtain the complex A-scan signal. Only the absolute values are  stored as the A-scan data. 

\subsubsection{IVOCT Volume Reconstruction}
In an IVOCT setup, the OCT device acquires a temporal sequence of A-scans during rotation and pullback. A sequence of $N$  A-scans represents one complete rotation of the prism, where $N$ depends on the acquisition rate and the rotation speed. Although the A-scans are actually acquired on a helical trajectory due to the continuous pullback, they are usually visualized as a planar slice. This is justified by the very small pullback distance $s_\text{rot} = v_\text{pull} / f_\text{rot} = \SI{45}{\micro\meter}$ during a rotation.

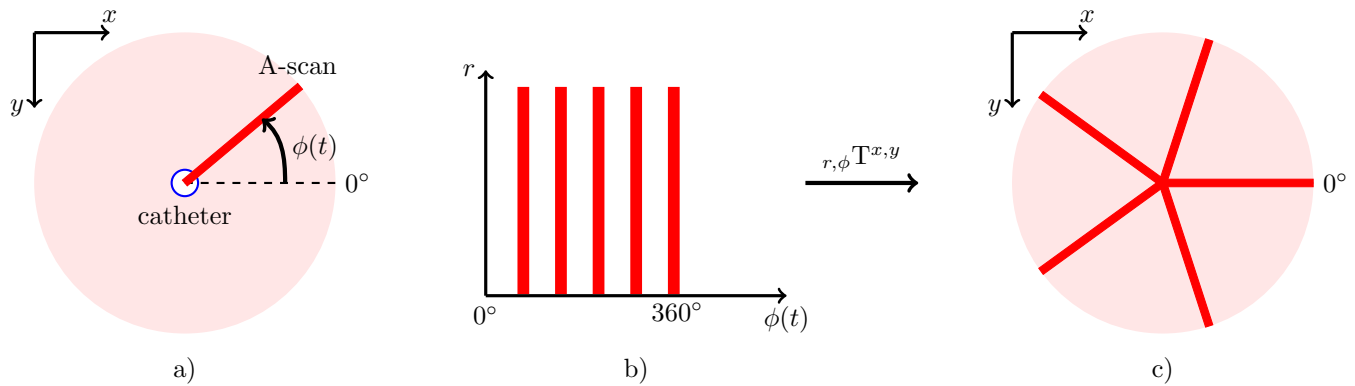
\begin{figure}[!tb] \centering
\begin{tikzpicture}
\draw[->,very thick] (-2,2) -- ++(1,0) node[above]{$x$};
\draw[->,very thick] (-2,2) -- ++(0,-1) node[left]{$y$};
\fill[red!10] (0,0) circle (57pt);
\draw[fill=white,draw=blue,thick] (0,0) circle (5pt) node[below,yshift=-5]{catheter};
\draw[dashed,thick] (0,0) -- (2,0) node[right]{\SI{0}{\degree}};
\draw[red,fill,rotate around={40:(0,0)}] (0,-0.05) rectangle (2,0.05) node[above,black]{A-scan};
\draw[->,ultra thick] (4/3,0) to[out=90,in=-40] node[midway,right]{$\phi(t)$} (4/3*0.77,4/3*0.64);

\draw[->,very thick](4.0,-1.5)node[below]{\SI{0}{\degree}} -- ++(4,0) node[below]{$\phi(t)$};
\draw[->,very thick](4.0,-1.5) -- ++(0,3) node[left,black]{$r$};
\foreach \i in {1,2,...,4}
	\draw[red,ultra thick, fill](4+0.05+\i*0.5,-1.45)rectangle (4-0.05+\i*0.5,1.25);
\draw[red,ultra thick, fill](6.55,-1.45) node[below,black]{\SI{360}{\degree}} rectangle (6.45,1.25);
\draw[->,ultra thick](8.25,0)--node[above]{$_{r,\phi}$T$^{x,y}$}++(1.5,0);

\draw[->,very thick] (11,2) -- ++(1,0) node[above]{$x$};
\draw[->,very thick] (11,2) -- ++(0,-1) node[left]{$y$};
\fill[red!10] (13,0) circle (57pt);
\draw[dashed,thick] (13,0) -- (15,0) node[right]{\SI{0}{\degree}};
\draw[red,fill,rotate around={0:(13,0)}] (13,-0.05) rectangle (15,0.05) node[above,black]{};
\draw[red,fill,rotate around={72:(13,0)}] (13,-0.05) rectangle (15,0.05) node[above,black]{};
\draw[red,fill,rotate around={144:(13,0)}] (13,-0.05) rectangle (15,0.05) node[above,black]{};
\draw[red,fill,rotate around={216:(13,0)}] (13,-0.05) rectangle (15,0.05) node[above,black]{};
\draw[red,fill,rotate around={288:(13,0)}] (13,-0.05) rectangle (15,0.05) node[above,black]{};

\node[black] at (0,-2.5){a)};
\node[black] at (6,-2.5){b)};
\node[black] at (13,-2.5){c)};
\end{tikzpicture}
\caption{a) IVOCT B-scan acquisition. By continuous rotation of the prism inside the catheter (blue circle), a sequence of radial A-scans (bright red line) at time-dependent angles $\phi(t)$ is acquired. b) A-scans covering a complete cycle (light red disk) are represented in polar coordinates. c) Respective A-scans in Cartesian coordinates. The actual cross-sectional IVOCT B-scan in Fig.~\ref{fig:Figure06} is related to the Cartesian representation.} 
\label{fig:Figure05}
\end{figure}

The slice is created by discretizing the interval $[0,2\pi)$ to $N$ equidistant points and assigning them to the A-scans. As each pixel of an A-scan represents a certain depth $r$, we can interpret the sequence as a polar representation of the slice as shown in Fig.~\ref{fig:Figure05}b). After transformation to Cartesian space, we obtain a cross-sectional image (B-scan) of the vessel. The acquisition is illustrated in Fig.~\ref{fig:Figure05}c) for the static case without pullback.

\begin{figure}[!bt]
\minipage[t]{0.45\textwidth}
\begin{tikzpicture}
\node[inner sep=0pt] (a) at (0,0) {
\includegraphics[width=1.0\linewidth]{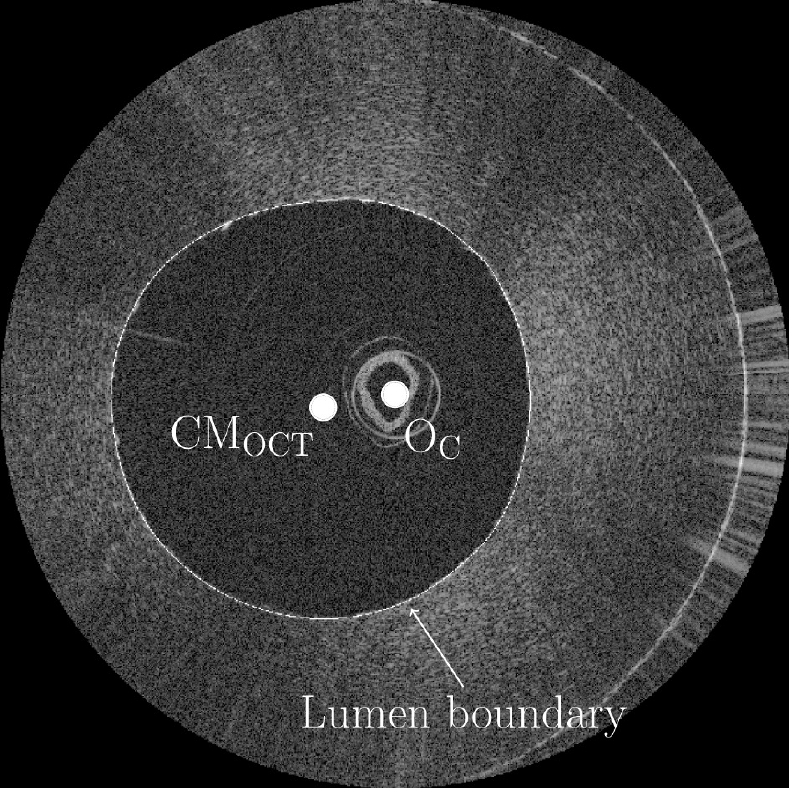}};
\end{tikzpicture}
\vspace{0.0pt}
\caption{The image origin $\text{O}_\text{C}$ of each B-scan is used for aligning in method A. Instead, methods B and C use the center of mass CM$_\text{OCT}$ of the lumen boundary.}
\label{fig:Figure06}
\endminipage\hfill
\minipage[t]{0.45\textwidth}
\includegraphics[width=1.0\linewidth]{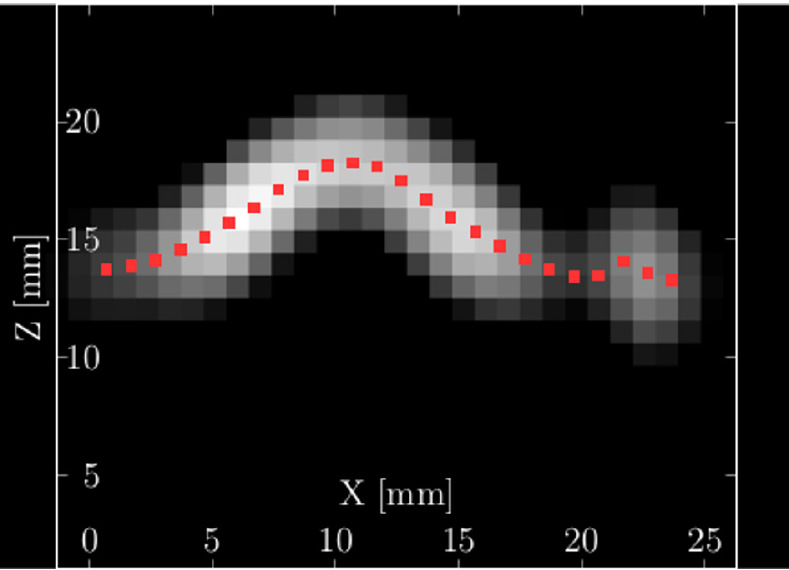}
\vspace{0.1pt}
\caption{Center of mass CM$_\text{MPI}$ in each MPI $yz$-slice (red squares) for a U-shaped vessel phantom shown in $xy$-slice. The trajectory is used to align the IVOCT B-scans to a volumetric image.}
\label{fig:Figure07}
\endminipage\
\end{figure}

To obtain a volumetric 3D image of the vessel phantom, we implemented several image processing steps in MATLAB which are described in the following. We first need to segment the lumen boundary in each B-scan (Fig.~\ref{fig:Figure06}). In order to simplify the segmentation process, the polar representation of the B-scans in Fig.~\ref{fig:Figure05}b) is used. The boundary is segmented with an edge detection algorithm after several preprocessing steps (removal of catheter, filtering, thresholding, morphological operations) as introduced by de Macedo et al.\cite{lumen2016}.
After lumen segmentation, we need to spatially align the acquired sequence of Cartesian B-scans and containing lumen areas. With the pullback distance $s$=\SI{30}{\milli\meter} and distance $l$=\SI{25}{\milli\meter} of the phantom landmarks, each IVOCT pullback consists of B-scans with narrowed lumen diameter. These landmarks are used to crop the IVOCT volumes on both sides and co-register the MPI and IVOCT data in method C to each other. This co-registration is applied with an implementation in MATLAB. Additionally, for morphologically correct reconstruction, the spatial trajectory of the catheter within the vessel has to be estimated. Even for a straight vessel, it will not be perfectly straight in practice because the catheter tip is subject to motion, e.g., due to pulsation, blood flow, or curvature in the path to the adapter. Tip motion estimation is further complicated by other artifacts, e.g., non-uniform rotations (NURD) \cite{tearney2012}. We will consider three different methods to align B-scans, two state-of-the-art reference methods labeled A and B and our proposed MPI-based pathway estimation method C. 

\begin{figure}[!bt] \centering
\begin{tabular}{ccc}		
 Method A & Method B & Method C \\
 \includegraphics[width=0.3\linewidth]{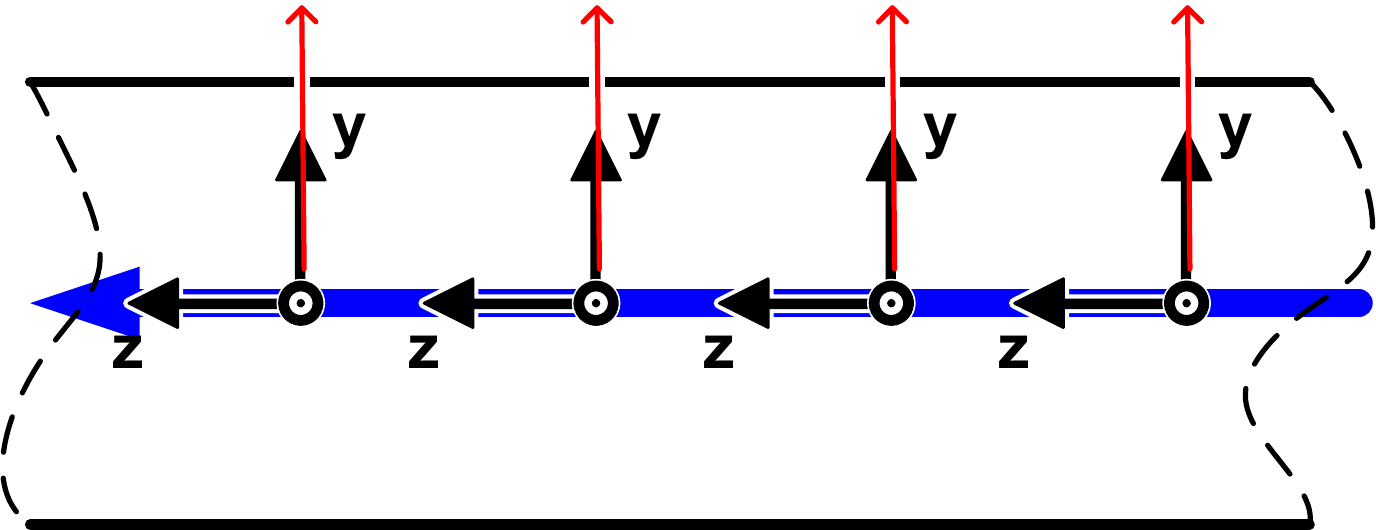} &
 \includegraphics[width=0.3\linewidth]{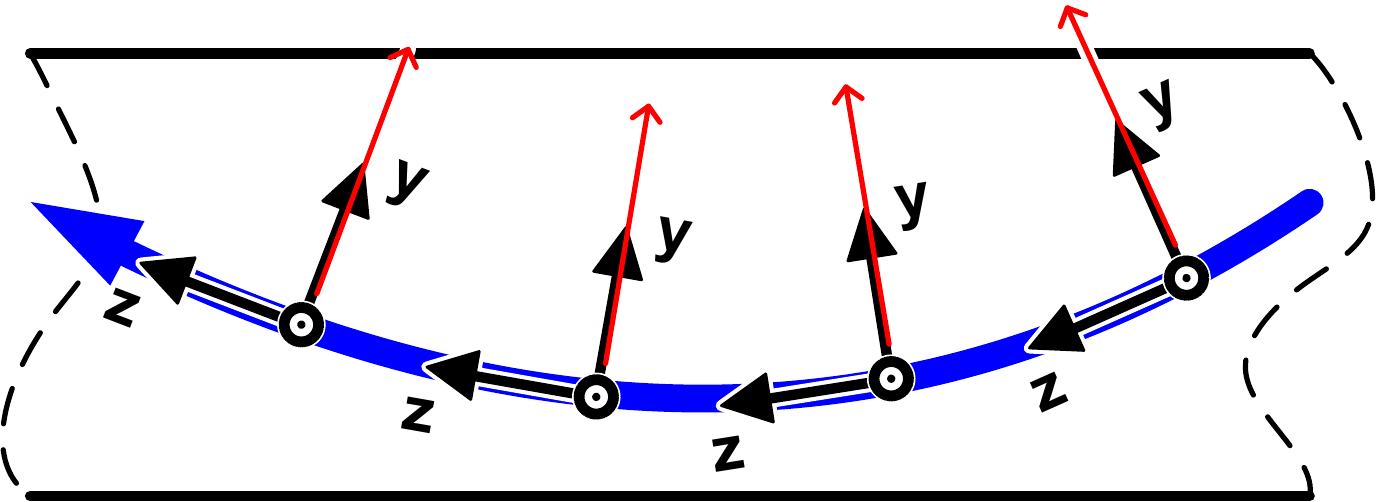} &
 \includegraphics[width=0.3\linewidth]{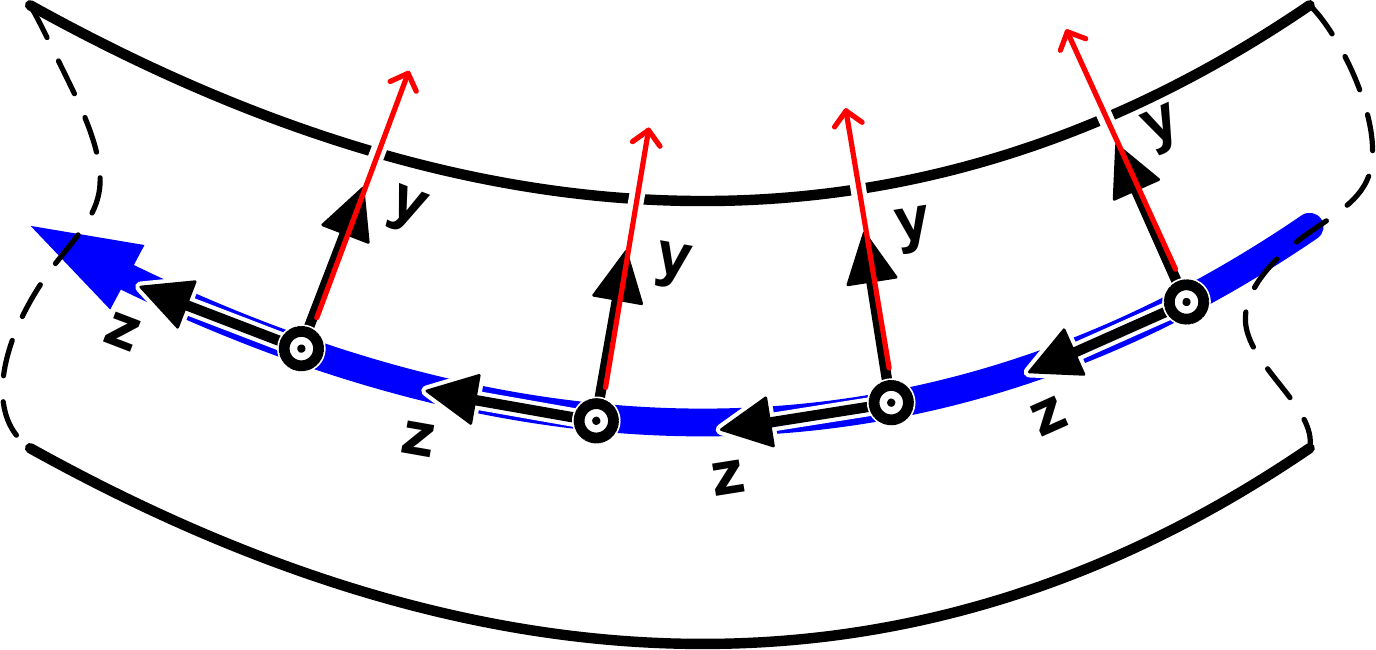}
\end{tabular}
\caption{Assumptions and limitations of the three investigated IVOCT volume reconstruction methods. Left: Method A stacks the origins of the B-scans, i.e., catheter centers, along a line and is able to reconstruct a straight vessel if the catheter trajectory is parallel to it. Center: Instead, method B uses the center of mass of the vessel lumen in each image as the reference point. Thus, it is possible to recover a straight vessel shape, even if the catheter is not at the same position in each cross-sectional image. Right: In method C, the centerline of the vessel is extracted from MPI data and used to estimate the catheter trajectory. Using this method, arbitrary vessel shapes can be recovered in the reconstructed IVOCT volume.}
\label{fig:Figure08}
\end{figure}

\paragraph*{Method A}
In the simplest approach, the origins O$_\text{c}$ (Fig.~\ref{fig:Figure06}) of the B-scans are equidistantly stacked along a straight line of length $l$. This approach assumes a perfectly straight catheter trajectory and neglects any motion of the catheter as well as the structure of the vessel (Fig.~\ref{fig:Figure08}, left). 

\paragraph*{Method B}
The previously segmented lumen boundary is used to calculate the lumen's center of mass CM$_\text{OCT}$ (Fig.~\ref{fig:Figure06}) \cite{Athanasiou2012}. Afterwards, the B-scans are equidistantly stacked such that their CM$_\text{OCT}$ are located on a straight line. This method takes into account that the catheter will not always be at the same position within the vessel cross-section (Fig.~\ref{fig:Figure08}, center).

\paragraph*{Method C}
The method also uses the center of mass of the lumen for aligning the B-scans. But instead of assuming a straight line, the trajectory of the catheter is estimated from MPI data. The static 3D MPI tomogram is analyzed slice-wise along the $x$-direction using our software tool written in Julia. For each $yz$-slice, the center of mass CM$_{\text{MPI}}$ is determined using a submillimeter-accurate algorithm \cite{griese2017submillimeter}. By interpolation of the positions CM$_{\text{MPI}}$ to the number of IVOCT B-scans, the corrected trajectory follows. The CM$_\text{OCT}$ of each B-scan is then placed along the resulting trajectory (Fig.~\ref{fig:Figure08}, right) reflecting the actual shape of the vessel as depicted in Fig.~\ref{fig:Figure07}.

After the spatial alignment of the B-scans, the segmented lumen boundaries are used to generate the 3D image volume. 

\subsection{Experiments}

The experimental approaches to study the trade-off with respect to the signal strength and the feasibility of bimodal imaging and image reconstruction are described below. Additionally, an \textit{in-vivo} feasibility study to estimate the shape of vessels in a mouse model is illustrated.

\subsubsection{Influence of SPIOs on the optical signal quality}
To investigate scattering, absorption and reflection in the OCT data related to the MPI sensitive SPIOs, we considered a series of different perimag SPIO concentrations as listed in Table~\ref{tab:Table1}. All concentrations are detectable with the MPI device whereas the image quality in terms of SNR and spatial resolution improves with the SPIO concentration \cite{Graeser2017}. 

Each solution from the series is filled inside a vessel phantom with an inner diameter of \SI{2.5}{\milli\meter} and an OCT measurement is performed using only the rotation and no pullback. The OCT data is analyzed with respect to the OCT mean signal intensity around the vessel border. With the help of optical assessment and the course of the mean signal intensity a suitable SPIO concentration is determined.

\begin{table}[hbt]
\centering
\caption{Concentrations of the dilution series used for optical fraction measurements. Water without perimag SPIOs is used as reference concentration $c_{10}$.}
 \label{tab:Table1}
 \begin{tabular}{ l  c  c  c  c  c  c  c  c  c  c }
 \hline \hline
   num & $c_1$ & $c_2$ & $c_3$ & $c_4$ & $c_5$ & $c_6$ & $c_7$ & $c_8$ & $c_9$ & $c_{10}$ \\ \hline
   dilution [n] & 1 & 2 & 4 & 8 & 16 & 32 & 64 & 128 & 256   & $-$ \\ 
   iron concentration [\SI{}{\milli\mol\per\liter}] & 50 & 25 & 12.5 & 6.75 & 3.13 & 1.56 & 0.78 & 0.39 & 0.19  & 0\\ \hline \hline
 \end{tabular}
\end{table}

\subsubsection{Phantom Study}
To study the actual volume reconstruction, MPI images and A-scans were acquired for the sequential and simultaneous acquisition scenarios. We use three differently shaped phantoms (stenosis-like, Z-shaped, and U-shaped) shown in Fig.~\ref{fig:Figure04}.

\subsubsection{In-vivo Feasibility Study}
To investigate the feasibility to perform simultaneous measurements of MPI and IVOCT, we have conducted an \textit{in-vivo} MPI experiment in a mouse where \SI{60}{\micro\liter} Resovist has been injected as a blood pool tracer in the tail vein of an anesthetized mouse. The injected bolus of iron oxide particles has a concentration of 87 mmol(Fe)/L. The MPI selection field gradient of \SI{1.5}{\tesla\per\meter} with an excitation amplitude of \SI{14}{\milli\tesla} is chosen. This results in MPI FoV of \SI{37x37x19}{\milli\meter}.
In order to obtain anatomical background information of the \textit{in-vivo} mouse, an Magnetic Resonance Imaging (MRI) scan is performed previously before the MPI experiment. The MRI scan is conducted with a 7 T preclinical MR scanner (Bruker) using a respiratory triggered T$_2$-weighted 2D turbo spin echo sequence in sagittal orientation. The chosen MRI parameters for the 2D turbo spin echo sequence are a FoV of \SI{44.6x90.0}{\milli\meter} with matrix discretization of 222 x 448 and 28 slices with a thickness 0.8mm. Both images are registered manually by rigid transformations with the help of fiducials, which can be identified in both imaging modalities.

\section{Results}
\subsection{Influence of SPIOs on the optical signal quality}
\begin{figure*}[!tb]
\minipage{1.0\textwidth}
\includegraphics[width=1.0\linewidth]{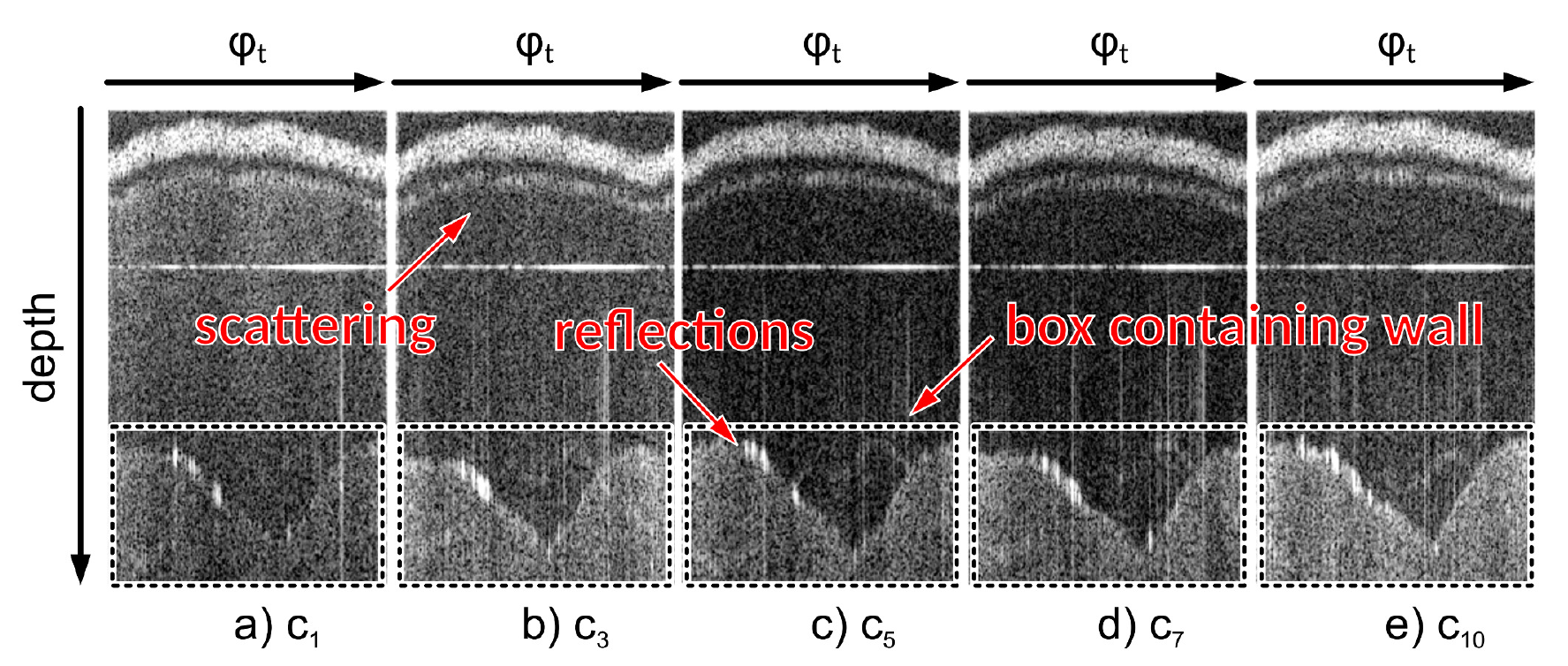}
\caption{Polar IVOCT images for perimag concentrations c$_1$, c$_{3}$, c$_{5}$, c$_{7}$, and c$_{10}$ (water) inside the phantom. At the highest concentration c$_1$ in a) you can identify a slightly denser scattering as compared to the other concentrations. This effect is underlined by a less intense signal at the edge of the phantom compared to the other concentration.}
\label{fig:Figure09}
\endminipage
\end{figure*}

\begin{figure}[!tb]
\centering
\includegraphics[width=0.4\linewidth]{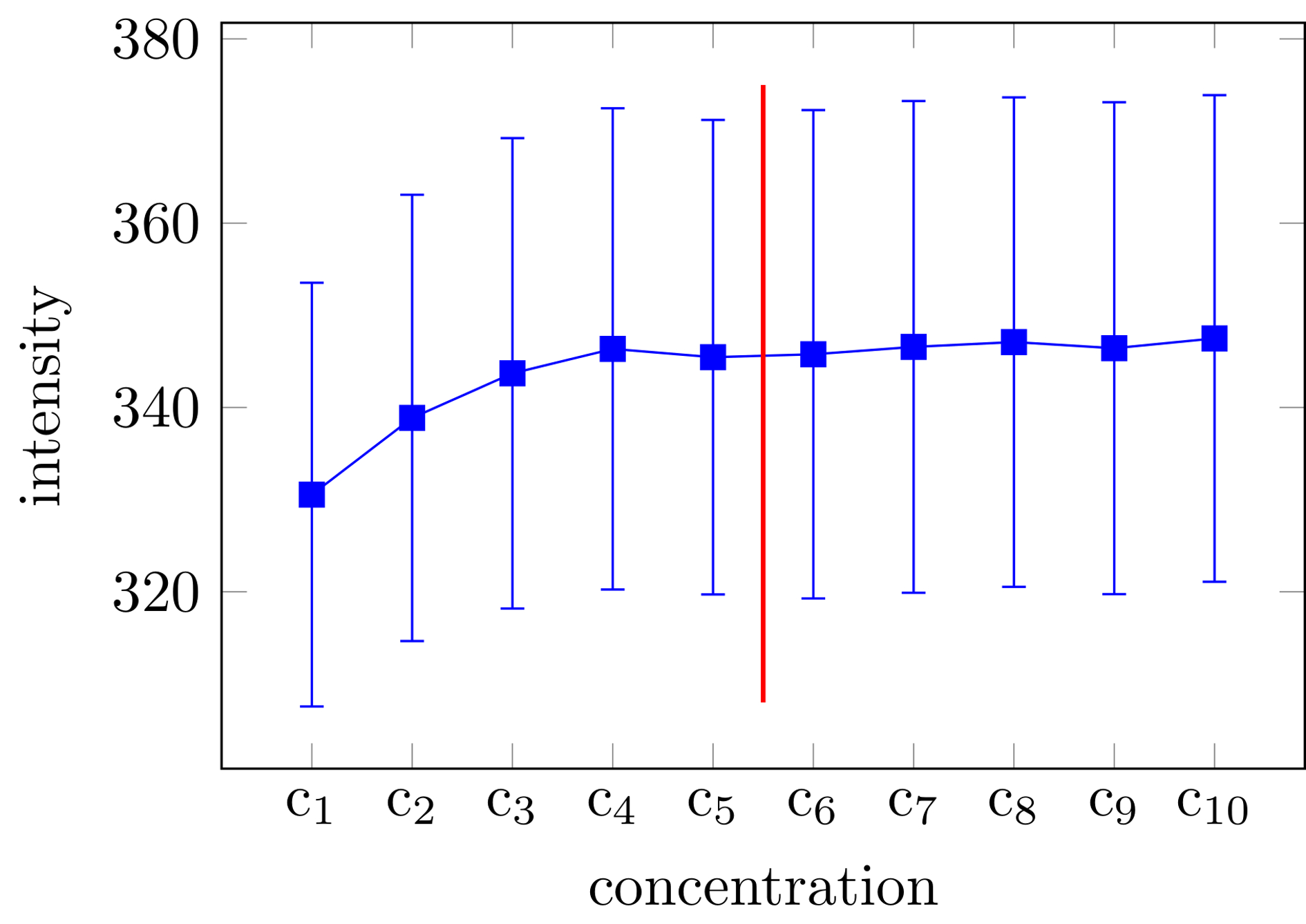}
\caption{Mean intensities with standard deviations from area of the vessel walls (dashed black box in Fig.~\ref{fig:Figure09}) for perimag dilution series $c_1$-$c_{10}$. The mean intensity increases up to the concentration of $c_4$ and remains on the same level for the following concentrations.}
\label{fig:Figure10}
\end{figure}

Normalized IVOCT images in polar coordinates for concentrations $c_1$, $c_3$, $c_5$, $c_7$, and $c_{10}$ are shown in Fig.~\ref{fig:Figure09}(a-e). For the highest concentration $c_1$ in Fig.~\ref{fig:Figure09}(a) more scattering is visible. This results in lower reflection intensity values in the vessel walls compared to vessel walls at concentration $c_3$, $c_5$, $c_7$, and $c_{10}$ in Fig.~\ref{fig:Figure09}(b-e). By comparing the OCT images of $c_5$-$c_{10}$, one can notice almost no difference with respect to scattering and reflection at the vessel wall. This visual impression agrees with a comparison of mean intensities in an area containing the vessel wall (dashed black box in Fig.~\ref{fig:Figure09}). They are plotted for all concentrations in Fig.~\ref{fig:Figure10} and show that the mean intensity does not increase after concentration $c_4$-$c_5$. Therefore, a higher dilution factor than n=$16$-$32$ does not substantially improve the OCT signal at the vessel wall.

Based on these findings, we selected a concentration of $c_{\text{exp}}=\SI{2.5}{\milli\mol\per\liter}$ with a dilution factor of $n=20$ for the pullback experiments. This concentration between concentrations $c_5$ and $c_6$ represents a suitable compromise between signal strength in OCT and MPI measurements, respectively. An influence on the refractive index of water (1.33) and the related OCT pixel spacing could not be identified (compare Section \ref{sec:IVOCTAcq}). 

\subsection{Phantom Study}
\subsubsection{Pathway Estimation}
\begin{figure}[!tb]
\begin{tabular}{ ccc}		
 Stenosis & Z-shape & U-shape \\
\includegraphics[width=0.31\linewidth]{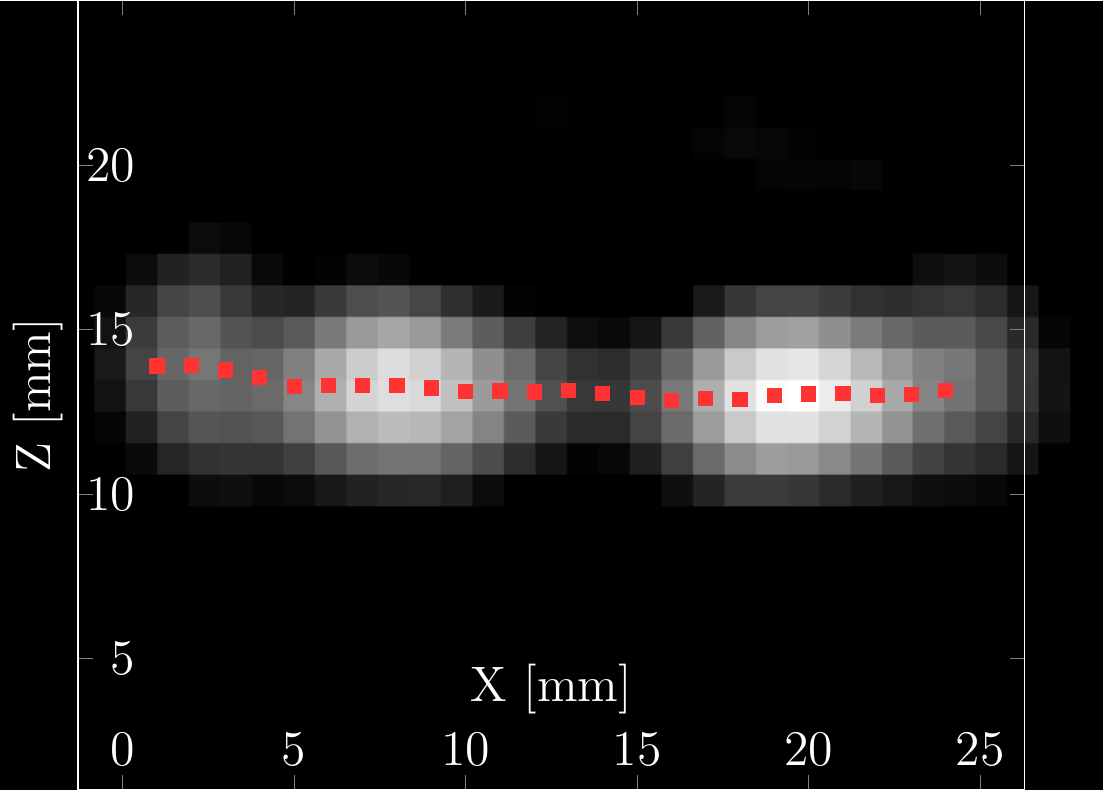} &
\includegraphics[width=0.31\linewidth]{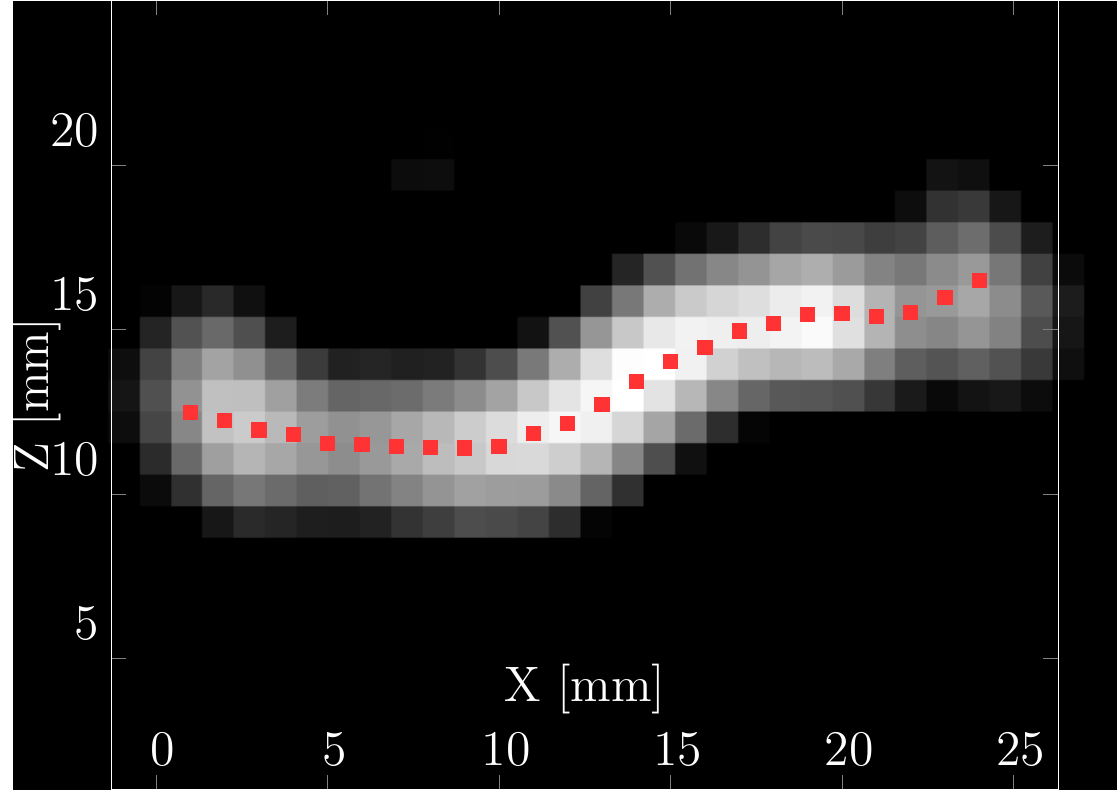} &
\includegraphics[width=0.31\linewidth]{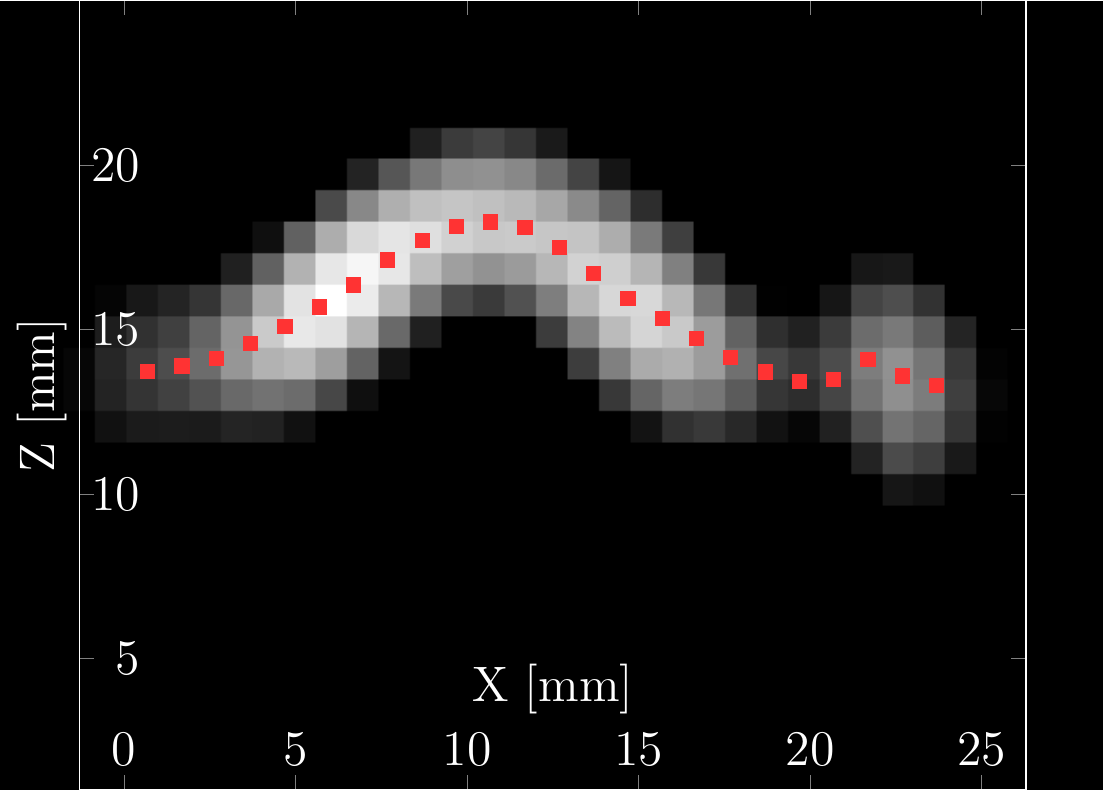}
\end{tabular}
\caption{Reconstructed MPI volumes as $xz$-slice of stenosis, Z-shape, and U-shape phantom. The red squares represent the center-of-mass CM$_\text{MPI}$ positions per $yz$-slice and result in the estimated three-dimensional IVOCT pullback pathway used for method C.}
\label{fig:Figure11}
\end{figure}

In Fig.~\ref{fig:Figure11}, the reconstructed 3D MPI volumes are shown as $xz$-slices for the stenosis, Z-shape and U-shape phantoms, respectively. The estimated center of mass CM$_\text{MPI}$ positions per $yz$-slice are depicted as red squares.

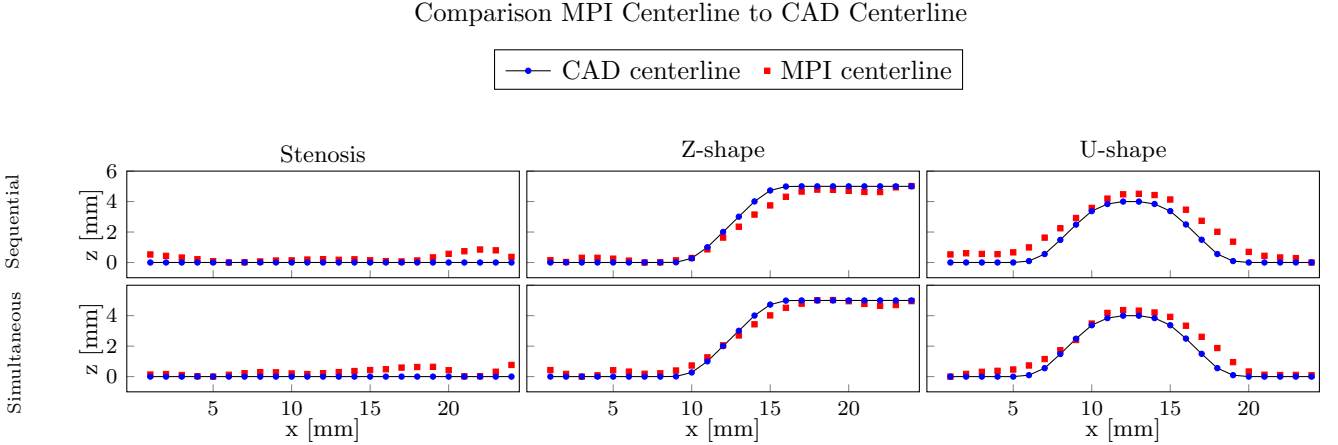
\begin{figure*}[!tb]
\begin{tikzpicture}
\begin{scope}[shift={(-6,-3.5)}]
\begin{groupplot}[
    group style={
        group name=CADOCTCenterline,
        group size=3 by 2,
        xlabels at=edge bottom,
        xticklabels at=edge bottom,
        vertical sep=1mm,
        horizontal sep=1mm,
        ylabels at=edge bottom,
    },
    footnotesize,
    width=6.8cm,
    height=3cm,
    xmin=0, xmax=25,
    ymin=1.4, ymax=3,
    xtick={5,10,15,20},
    xticklabels={5,10,15,20},
    tickpos=left,
    ytick={0,2,...,6},
    xticklabels={1.5,2.0,2.5},
    xlabel style={yshift=2mm},
    ylabel style={yshift=-2mm},
]
\nextgroupplot[ymin=-1,ymax=6, xmin=-0.5, xmax=24.5,ytick={0,2,...,6},xtick={5,10,15,20},
    ylabel={z [mm]}, title={Stenosis}]
\addplot+ [mark = {square*}, only marks = {true}, mark size = {1}, red,mark options={red}]coordinates {
(1.0, 0.5267372131347656)
(2.0, 0.4368009567260742)
(3.0, 0.324310302734375)
(4.0, 0.21321821212768555)
(5.0, 0.08269548416137695)
(6.0, 0.0)
(7.0, 0.014151573181152344)
(8.0, 0.06682443618774414)
(9.0, 0.12865400314331055)
(10.0, 0.13895130157470703)
(11.0, 0.1919693946838379)
(12.0, 0.21804237365722656)
(13.0, 0.17776966094970703)
(14.0, 0.20717811584472656)
(15.0, 0.14510107040405273)
(16.0, 0.09182024002075195)
(17.0, 0.06775951385498047)
(18.0, 0.1328134536743164)
(19.0, 0.3332042694091797)
(20.0, 0.5698328018188477)
(21.0, 0.7377147674560547)
(22.0, 0.8506960868835449)
(23.0, 0.7976069450378418)
(24.0, 0.3659343719482422)
};\label{plots:plot1}
\addplot+ [mark = {*}, mark size = {1}, black,mark options={blue}]coordinates {
(1, 0)
(2, 0)
(3, 0)
(4, 0)
(5, 0)
(6, 0)
(7, 0)
(8, 0)
(9, 0)
(10, 0)
(11, 0)
(12, 0)
(13, 0)
(14, 0)
(15, 0)
(16, 0)
(17, 0)
(18, 0)
(19, 0)
(20, 0)
(21, 0)
(22, 0)
(23, 0)
(24, 0)
};\label{plots:plot2}
\nextgroupplot[ymin=-1,ymax=6, xmin=-0.5, xmax=24.5, ytick={0,2,...,6},yticklabels={},xtick={5,10,15,20},, title={Z-shape}]
\addplot+ [mark = {square*}, only marks = {true}, mark size = {1}, red,mark options={red}]coordinates {
(1.0, 0.15005970001220703)
(2.0, 0.03415536880493164)
(3.0, 0.3105430603027344)
(4.0, 0.30985403060913086)
(5.0, 0.2507953643798828)
(6.0, 0.12228012084960938)
(7.0, 0.0)
(8.0, 0.012531757354736328)
(9.0, 0.1384439468383789)
(10.0, 0.2946295738220215)
(11.0, 0.8619933128356934)
(12.0, 1.635241985321045)
(13.0, 2.348987102508545)
(14.0, 3.1464762687683105)
(15.0, 3.747110366821289)
(16.0, 4.315835952758789)
(17.0, 4.65956974029541)
(18.0, 4.786473274230957)
(19.0, 4.771299362182617)
(20.0, 4.7046709060668945)
(21.0, 4.629844665527344)
(22.0, 4.624262809753418)
(23.0, 4.931845664978027)
(24.0, 5.022157669067383)
};
\addplot+ [mark = {*}, mark size = {1}, black,mark options={blue}]coordinates {
(1, 0)
(2, 0)
(3, 0)
(4, 0)
(5, 0)
(6, 0)
(7, 0)
(8, 0)
(9, 0.010880108247434794)
(10, 0.27223192776322724)
(11, 1.0017391304347827)
(12, 2.003913043478261)
(13, 3.006086956521739)
(14, 4.008260869565217)
(15, 4.727768072236772)
(16, 4.989119891752565)
(17, 5)
(18, 5)
(19, 5)
(20, 5)
(21, 5)
(22, 5)
(23, 5)
(24, 5)
};
\nextgroupplot[ymin=-1,ymax=6, xmin=-0.5, xmax=24.5,ytick={0,2,...,6},yticklabels={},xtick={5,10,15,20},, title={U-shape}]
\addplot+ [mark = {square*}, only marks = {true}, mark size = {1}, red,mark options={red}]coordinates {
(1.0, 0.5290336608886719)
(2.0, 0.6062264442443848)
(3.0, 0.5626654624938965)
(4.0, 0.544731616973877)
(5.0, 0.6645784378051758)
(6.0, 0.9906377792358398)
(7.0, 1.631718635559082)
(8.0, 2.2467966079711914)
(9.0, 2.9216079711914062)
(10.0, 3.584595203399658)
(11.0, 4.195310831069946)
(12.0, 4.476044178009033)
(13.0, 4.511301040649414)
(14.0, 4.428490877151489)
(15.0, 4.134909629821777)
(16.0, 3.4676361083984375)
(17.0, 2.738215923309326)
(18.0, 2.0171103477478027)
(19.0, 1.3674407005310059)
(20.0, 0.6965804100036621)
(21.0, 0.4402318000793457)
(22.0, 0.33010244369506836)
(23.0, 0.2881345748901367)
(24.0, 0.0)
};
\addplot+ [mark = {*}, mark size = {1}, black,mark options={blue}]coordinates {
(1.0, 0.0)
(2.0, 0.0)
(3.0, 0.0)
(4.0, 0.0)
(5.0, 0.0)
(6.0, 0.0938716374233266)
(7.0, 0.5541357081509757)
(8.0, 1.485217391304348)
(9.0, 2.487391304347827)
(10.0, 3.3735580686847864)
(11.0, 3.8416563547091482)
(12.0, 4.0)
(13.0, 4.0)
(14.0, 3.841656354709148)
(15.0, 3.3735580686847846)
(16.0, 2.497391304347826)
(17.0, 1.495217391304346)
(18.0, 0.5541357081509739)
(19.0, 0.09387163742332572)
(20.0, 0.0)
(21.0, 0.0)
(22.0, 0.0)
(23.0, 0.0)
(24.0, 0.0)
};
\nextgroupplot[ymin=-1,ymax=6, xmin=-0.5, xmax=24.5,ytick={0,2,...,4},xtick={5,10,15,20},xticklabels={5,10,15,20},xlabel={x [mm]},ylabel={z [mm]}]
\addplot+ [mark = {square*}, only marks = {true}, mark size = {1}, red,mark options={red}]coordinates {
(1.0, 0.13554906845092773)
(2.0, 0.16603326797485352)
(3.0, 0.10373401641845703)
(4.0, 0.02642345428466797)
(5.0, 0.0)
(6.0, 0.11950254440307617)
(7.0, 0.22296619415283203)
(8.0, 0.2904052734375)
(9.0, 0.27809667587280273)
(10.0, 0.20623207092285156)
(11.0, 0.16673660278320312)
(12.0, 0.22348737716674805)
(13.0, 0.29530906677246094)
(14.0, 0.3652215003967285)
(15.0, 0.4307279586791992)
(16.0, 0.48125314712524414)
(17.0, 0.591850757598877)
(18.0, 0.6288094520568848)
(19.0, 0.6403546333312988)
(20.0, 0.4178752899169922)
(21.0, 0.01925373077392578)
(22.0, 0.028218746185302734)
(23.0, 0.3208141326904297)
(24.0, 0.7653779983520508)
};
\addplot+ [mark = {*}, mark size = {1}, black,mark options={blue}]coordinates {
(1, 0)
(2, 0)
(3, 0)
(4, 0)
(5, 0)
(6, 0)
(7, 0)
(8, 0)
(9, 0)
(10, 0)
(11, 0)
(12, 0)
(13, 0)
(14, 0)
(15, 0)
(16, 0)
(17, 0)
(18, 0)
(19, 0)
(20, 0)
(21, 0)
(22, 0)
(23, 0)
(24, 0)
};
\nextgroupplot[ymin=-1,ymax=6, xmin=-0.5, xmax=24.5,ytick={0,2,...,6},yticklabels={},yticklabels={},xtick={5,10,15,20},xticklabels={5,10,15,20},xlabel={x [mm]}]
\addplot+ [mark = {square*}, only marks = {true}, mark size = {1}, red,mark options={red}]coordinates {
(1.0, 0.4281139373779297)
(2.0, 0.17569947242736816)
(3.0, 0.0)
(4.0, 0.08202004432678223)
(5.0, 0.41736364364624023)
(6.0, 0.32661938667297363)
(7.0, 0.1868298053741455)
(8.0, 0.216170072555542)
(9.0, 0.40033459663391113)
(10.0, 0.7272472381591797)
(11.0, 1.2623770236968994)
(12.0, 2.0515925884246826)
(13.0, 2.701810121536255)
(14.0, 3.4340627193450928)
(15.0, 4.017890214920044)
(16.0, 4.519172430038452)
(17.0, 4.7902772426605225)
(18.0, 5.015997648239136)
(19.0, 5.028314828872681)
(20.0, 4.950102090835571)
(21.0, 4.773106336593628)
(22.0, 4.647083044052124)
(23.0, 4.699005365371704)
(24.0, 4.95503306388855)
};
\addplot+ [mark = {*}, mark size = {1}, black,mark options={blue}]coordinates {
(1, 0)
(2, 0)
(3, 0)
(4, 0)
(5, 0)
(6, 0)
(7, 0)
(8, 0)
(9, 0.010880108247434794)
(10, 0.27223192776322724)
(11, 1.0017391304347827)
(12, 2.003913043478261)
(13, 3.006086956521739)
(14, 4.008260869565217)
(15, 4.727768072236772)
(16, 4.989119891752565)
(17, 5)
(18, 5)
(19, 5)
(20, 5)
(21, 5)
(22, 5)
(23, 5)
(24, 5)
};
\nextgroupplot[ymin=-1,ymax=6, xmin=-0.5, xmax=24.5,ytick={0,2,...,6},yticklabels={},yticklabels={},xtick={5,10,15,20},xticklabels={5,10,15,20},xlabel={x [mm]}]
\addplot+ [mark = {square*}, only marks = {true}, mark size = {1}, red,mark options={red}]coordinates {
(1.0, 0.0)
(2.0, 0.17787599563598633)
(3.0, 0.31522274017333984)
(4.0, 0.380490779876709)
(5.0, 0.46510982513427734)
(6.0, 0.729771614074707)
(7.0, 1.152554988861084)
(8.0, 1.7287750244140625)
(9.0, 2.4024648666381836)
(10.0, 3.4821524620056152)
(11.0, 4.183754920959473)
(12.0, 4.367507457733154)
(13.0, 4.330708742141724)
(14.0, 4.2089197635650635)
(15.0, 3.922368288040161)
(16.0, 3.341740369796753)
(17.0, 2.6119680404663086)
(18.0, 1.8747749328613281)
(19.0, 0.9509162902832031)
(20.0, 0.3333554267883301)
(21.0, 0.1226348876953125)
(22.0, 0.10680770874023438)
(23.0, 0.11132192611694336)
(24.0, 0.08671092987060547)
};
\addplot+ [mark = {*}, mark size = {1}, black,mark options={blue}]coordinates {
(1.0, 0.0)
(2.0, 0.0)
(3.0, 0.0)
(4.0, 0.0)
(5.0, 0.0)
(6.0, 0.0938716374233266)
(7.0, 0.5541357081509757)
(8.0, 1.485217391304348)
(9.0, 2.487391304347827)
(10.0, 3.3735580686847864)
(11.0, 3.8416563547091482)
(12.0, 4.0)
(13.0, 4.0)
(14.0, 3.841656354709148)
(15.0, 3.3735580686847846)
(16.0, 2.497391304347826)
(17.0, 1.495217391304346)
(18.0, 0.5541357081509739)
(19.0, 0.09387163742332572)
(20.0, 0.0)
(21.0, 0.0)
(22.0, 0.0)
(23.0, 0.0)
(24.0, 0.0)
};
\end{groupplot}
\node[rotate=90] (s1) at (-1.5,0.75){\scriptsize Sequential};
\node[rotate=90] (s2) at (-1.5,-1.0){\scriptsize Simultaneous};

\end{scope};
\node[xshift=15mm] (s3){Comparison MPI Centerline to CAD Centerline};
\matrix[
    matrix of nodes,
    anchor=south,
    draw,
    inner sep=0.2em,
    draw
  ] at (2,-1) []
  {\ref{plots:plot2}& CAD centerline &[5pt]
    \ref{plots:plot1}& MPI centerline\\};
    %https://tex.stackexchange.com/questions/192424/pgfplots-single-legend-in-a-group-plot Lösung für Legende
\end{tikzpicture}
\caption{Estimated MPI catheter pathway compared to the CAD centerline for both scenarios and all three phantoms stenosis, Z-shape, and U-shape. The 3D centerlines are only depicted in $xz$-planes where the blue lines represent the centerlines of the CAD models and the red dots illustrate the center of mass CM$_\text{MPI}$ positions per $yz$-slice. These MPI centerlines are used for the volume reconstruction method C.}
\label{fig:Figure12}
\end{figure*}

In Fig.~\ref{fig:Figure12}, the estimated pathways from Fig.~\ref{fig:Figure11} are analyzed quantitatively by comparison to the centerline ground truth of the CAD models. The estimated pathways are given for the sequential and simultaneous application scenario and they are in good agreement with the centerlines from the CAD models for both scenarios. The mean absolute error (MAE) for both scenarios and all three phantoms is in the range of \SI{0.25}{\milli\meter} to \SI{0.28}{\milli\meter} as summarized in Table~\ref{tab:Table2}. The standard deviation $\sigma$ ranges from \SI{0.21}{\milli\meter} to \SI{0.35}{\milli\meter} for both scenarios and all phantoms. The paired t-test is used to calculate p-values for the differences of both methodologies sequential and simultaneous per phantom type. For the stenosis phantom the p-value is 0.95, for the Z-shape it is 0.57 and for the U-shape phantom the p-value is 0.55. It can be noted that the estimated MPI pathway is overall slightly flattened compared to the CAD centerline.

\def\distFigX{5.3}
\def\distFigY{-2.6}
\def\offsetY{-0.8}
\def\factorLW{0.31}
\def\sOffset{-3.0}

\begin{table}[!tb]
\centering
\caption{Mean absolute error (MAE) and standard deviation $\sigma$ calculated for the MPI estimated pathway to CAD centerline. We use a paired t-test to check the results of the two methods for a significant deviation.}
\label{tab:Table2}
\begin{tabular}{ l  l  c  c  c }
\hline \hline
  Phantoms & & Stenosis & Z-shape & U-shape  \\ \hline
  Sequential & MAE [\SI{}{\milli\meter}]  & 0.28 & 0.28 & 0.27 \\ 
  Sequential & $\sigma$ [\SI{}{\milli\meter}]  & 0.24 & 0.35 & 0.35 \\   
  Simultaneous & MAE [\SI{}{\milli\meter}]  & 0.28 & 0.26 & 0.25 \\ 
  Simultaneous & $\sigma$  [\SI{}{\milli\meter}]  & 0.21 & 0.32 & 0.34 \\
  p-value (paired t-test) & & 0.95 & 0.57 & 0.55 \\ \hline \hline
\end{tabular}
\end{table}

\subsubsection{Volume Reconstruction}
\begin{figure}[!tb]
\includegraphics[width=1.0\linewidth]{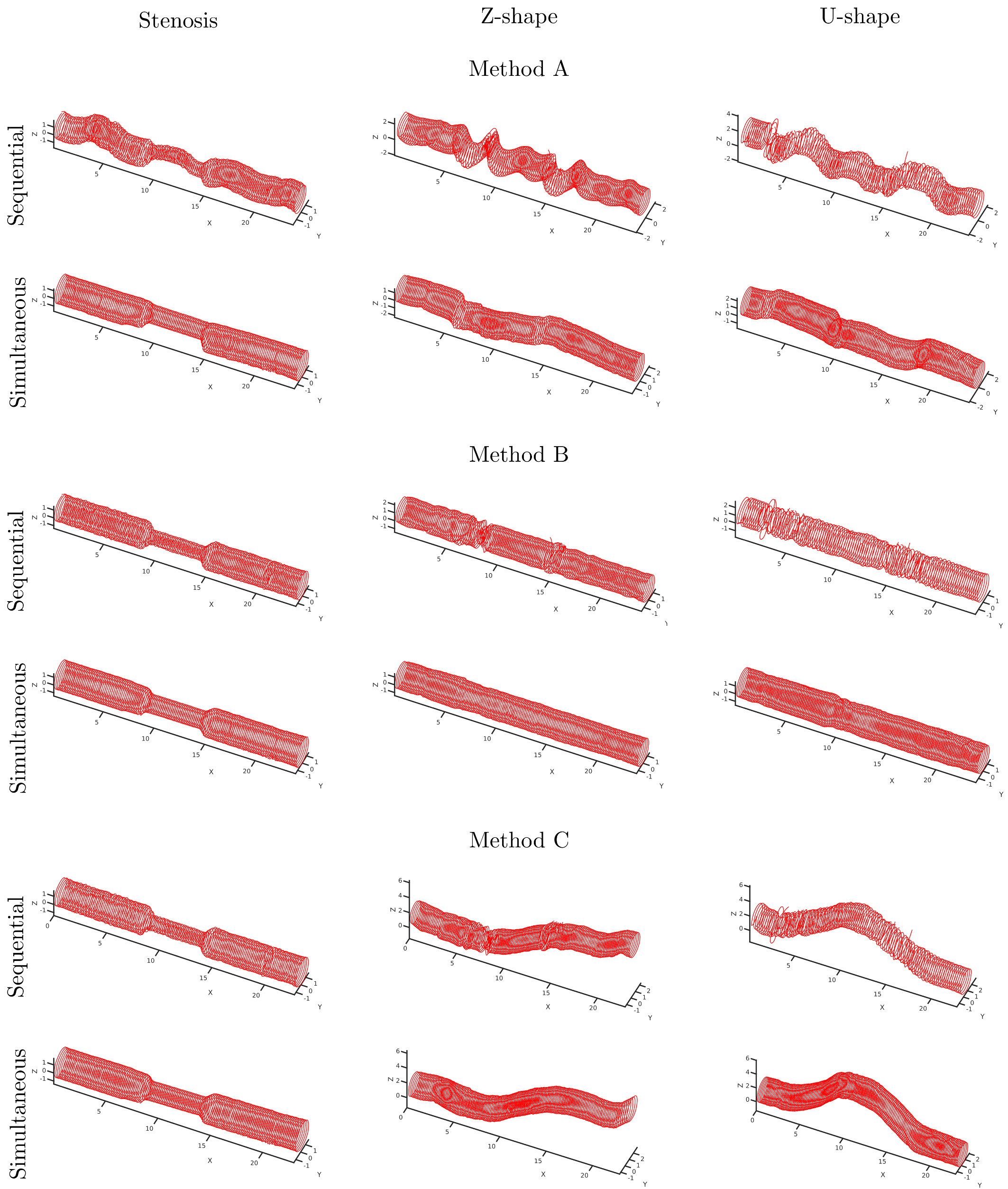}
\caption{IVOCT volume reconstructions for the three phantoms (stenosis, Z-shape, U-shape) of method A, B, and C shown for both scenarios.}
\label{fig:Figure13}
\end{figure}

The reconstructed phantom volumes based on the three reconstruction methods are shown in Fig.~\ref{fig:Figure13}.
\paragraph*{Method A:}
The reconstructed phantom IVOCT volumes for both methodologies and all three phantoms show high deviations from the CAD models. Twists and bends alongside the volume appear. In case of the stenosis phantom, one can identify the wide and narrow parts but for the Z-shape and U-shape phantom, one cannot detect their actual shape.
\paragraph*{Method B:}
The volumes reconstructed with method B provide a better representation of the phantom shape. In case of the stenosis, it resembles the phantom very much in both scenarios. In case of the Z-shape and U-shape phantom, the volumes appear as well shaped cylinders for both scenarios. But they still do not comply with the CAD models.
\paragraph*{Method C:}
For method C, the reconstructed volumes resemble the shape of all three phantoms quite well. For the stenosis phantom, the volumes for method C and method B look similar. For the Z-shape and U-shape volumes a clear difference is visible. Both reconstructed volume shapes are conform with the ground-truth shapes of their phantoms. 

\begin{figure*}[!tb]
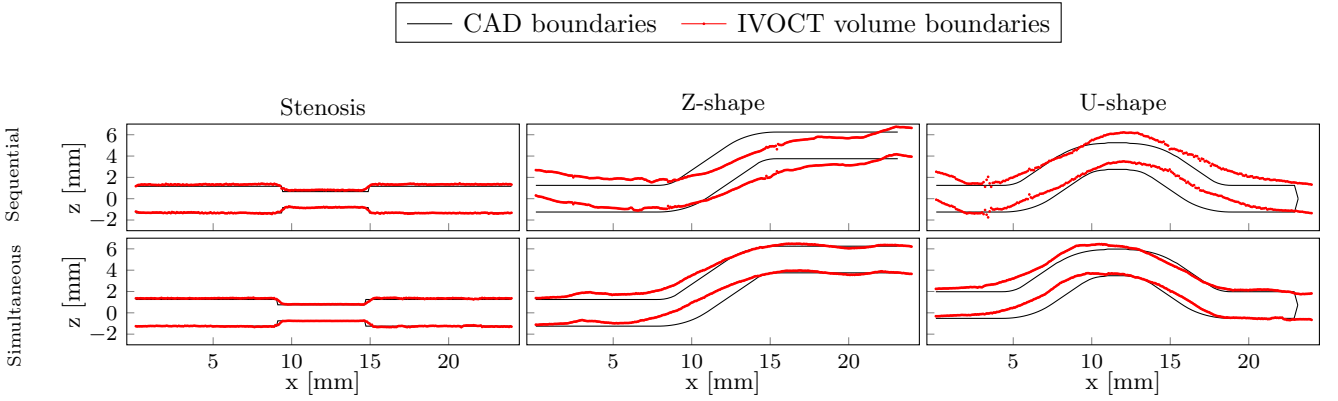

\begin{tikzpicture}
\begin{scope}[shift={(-6,-3.5)}]
\begin{groupplot}[
    group style={
        group name=my plots,
        group size=3 by 2,
        xlabels at=edge bottom,
        xticklabels at=edge bottom,
        vertical sep=1mm,
        horizontal sep=1mm,
        ylabels at=edge bottom,
    },
    footnotesize,
    width=6.8cm,
    height=3cm,
    xmin=0, xmax=25,
    ymin=-5, ymax=5,
    xtick={5,10,15,20},
    xticklabels={5,10,15,20},
    tickpos=left,
    ytick={1.5,2.0,2.5},
    xticklabels={1.5,2.0,2.5},
    xlabel style={yshift=2mm},
    ylabel style={yshift=-2mm},
]
\nextgroupplot[ymin=-3,ymax=7, xmin=-0.5, xmax=24.5,ytick={-2,0,...,6},xtick={5,10,15,20},
    ylabel={z [mm]}, title={Stenosis}]
\input{coordsStenose2D.tikz}
\nextgroupplot[ymin=-3,ymax=7, xmin=-0.5, xmax=24.5, ytick={-2,0,...,6},yticklabels={},xtick={5,10,15,20},, title={Z-shape}]
\input{coordsZShape2D.tikz}
\nextgroupplot[ymin=-3,ymax=7, xmin=-0.5, xmax=24.5,ytick={-2,0,...,6},yticklabels={},xtick={5,10,15,20},, title={U-shape}]
\input{coordsUShape2D.tikz}
\nextgroupplot[ymin=-3,ymax=7, xmin=-0.5, xmax=24.5,ytick={-2,0,...,6},xtick={5,10,15,20},xticklabels={5,10,15,20},xlabel={x [mm]},ylabel={z [mm]}]
\input{coordsScenario2_Stenose2D.tikz}
\nextgroupplot[ymin=-3,ymax=7, xmin=-0.5, xmax=24.5,ytick={-2,0,...,6},yticklabels={},yticklabels={},xtick={5,10,15,20},xticklabels={5,10,15,20},xlabel={x [mm]}]
\input{coordsScenario2_ZShape2D.tikz}
\nextgroupplot[ymin=-3,ymax=7, xmin=-0.5, xmax=24.5,ytick={-2,0,...,6},yticklabels={},yticklabels={},xtick={5,10,15,20},xticklabels={5,10,15,20},xlabel={x [mm]}]
\input{coordsScenario2_UShape2D.tikz}
\end{groupplot}
\node[rotate=90] (s1) at (-1.5,0.75){\scriptsize Sequential};
\node[rotate=90] (s2) at (-1.5,-1.0){\scriptsize Simultaneous};

\end{scope};
\node[xshift=18mm] (s3){2D IVOCT $xz$-view and CAD phantom borders};
\matrix[
    matrix of nodes,
    anchor=south,
    draw,
    inner sep=0.2em,
    draw
  ] at (2,-1) []
  {\ref{plots:plot4}& CAD boundaries &[5pt]
    \ref{plots:plot3}& IVOCT volume boundaries\\};
\end{tikzpicture}
\caption{$xz$-slice of the lumen boundaries of the IVOCT volume reconstructions for the stenosis, Z-shape and U-shape phantom in red and the CAD borders in black for both scenarios.}
\label{fig:Figure14}
\end{figure*}

The reconstructed volumes from method C are also shown as $xz$-slices in 2D together with the cross-sections of the CAD sketches in Fig.~\ref{fig:Figure14}. The OCT lumen boundary is shown in red while the CAD model borders are depicted in black. For both IVOCT volumes of the stenosis (simultaneous and sequential), the lumen boundaries largely coincide with the border of the phantoms. For the IVOCT volumes of the Z-shape phantom, the deviations from the CAD model increase and lumen boundaries appear a little more flattened. The lumen boundaries for the U-shape phantom show some deviation from the CAD model but generally they follow the phantom shape rather well for both scenarios.

\begin{figure*}[!tb]
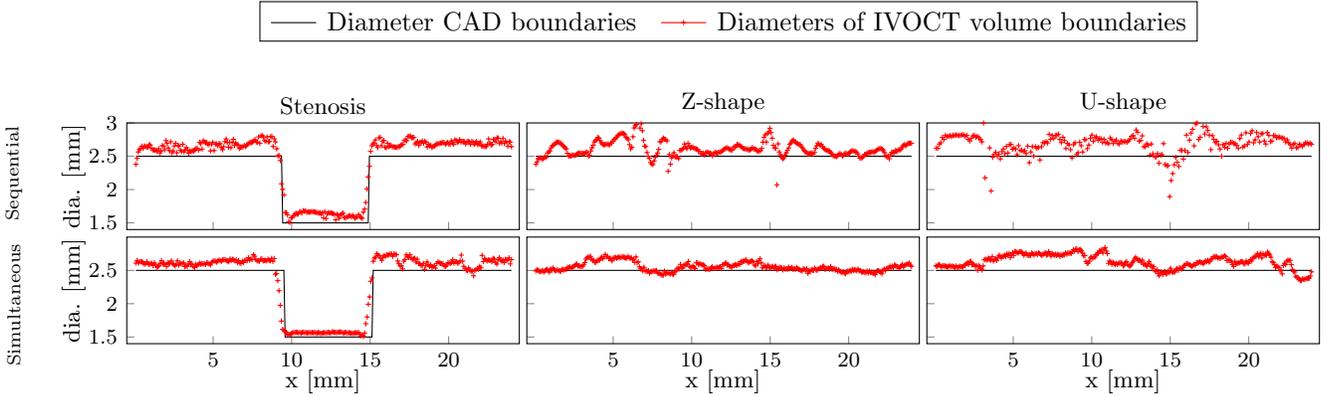

\begin{tikzpicture}
\begin{scope}[shift={(-6,-3.5)}]
\begin{groupplot}[
    group style={
        group name=my plots,
        group size=3 by 2,
        xlabels at=edge bottom,
        xticklabels at=edge bottom,
        vertical sep=1mm,
        horizontal sep=1mm,
        ylabels at=edge bottom,
    },
    footnotesize,
    width=6.8cm,
    height=3cm,
    xmin=0, xmax=25,
    ymin=1.4, ymax=3,
    xtick={5,10,15,20},
    xticklabels={5,10,15,20},
    tickpos=left,
    ytick={1.5,2.0,2.5},
    xticklabels={1.5,2.0,2.5},
    xlabel style={yshift=2mm},
    ylabel style={yshift=-2mm},
]
\nextgroupplot[ymin=1.4,ymax=3, xmin=-0.5, xmax=24.5,ytick={1.5,2.0,2.5,3.0},xtick={5,10,15,20},
    ylabel={dia. [mm]}, title={Stenosis}]
\input{coordsStenoseDiameter2D.tikz}
\nextgroupplot[ymin=1.4,ymax=3, xmin=-0.5, xmax=24.5, ytick={1.5,2.0,2.5},yticklabels={},xtick={5,10,15,20},, title={Z-shape}]
\input{coordsZShapeDiameter2D.tikz}
\nextgroupplot[ymin=1.4,ymax=3, xmin=-0.5, xmax=24.5,ytick={1.5,2.0,2.5},yticklabels={},xtick={5,10,15,20},, title={U-shape}]
\input{coordsUShapeDiameter2D.tikz}
\nextgroupplot[ymin=1.4,ymax=3, xmin=-0.5, xmax=24.5,ytick={1.5,2.0,2.5},xtick={5,10,15,20},xticklabels={5,10,15,20},xlabel={x [mm]},ylabel={dia. [mm]}]
\input{coordsScenarios2_StenoseDiameter2D.tikz}
\nextgroupplot[ymin=1.4,ymax=3, xmin=-0.5, xmax=24.5,ytick={1.5,2.0,2.5},yticklabels={},yticklabels={},xtick={5,10,15,20},xticklabels={5,10,15,20},xlabel={x [mm]}]
\input{coordsScenario2_ZShapeDiameter2D.tikz}
\nextgroupplot[ymin=1.4,ymax=3, xmin=-0.5, xmax=24.5,ytick={1.5,2.0,2.5},yticklabels={},yticklabels={},xtick={5,10,15,20},xticklabels={5,10,15,20},xlabel={x [mm]}]
\input{coordsScenario2_UShapeDiameter2D.tikz}
\end{groupplot}
\node[rotate=90] (s1) at (-1.5,0.75){\scriptsize Sequential};%-1.5
\node[rotate=90] (s2) at (-1.5,-1.0){\scriptsize Simultaneous};%-1.5
\end{scope};

\node[xshift=15mm] (s3){Comparison IVOCT diameter to CAD diameter};
\matrix[
    matrix of nodes,
    anchor=south,
    draw,
    inner sep=0.2em,
    draw
  ] at (2,-1) []
  {\ref{plots:plot6}& Diameter CAD boundaries &[5pt]
    \ref{plots:plot5}& Diameters of IVOCT volume boundaries\\};
\end{tikzpicture}
\caption{The diameters of the IVOCT volume boundaries compared with the three phantom diameters for the stenosis, Z-shape and U-shape phantoms given for both scenarios.}
\label{fig:Figure15}
\end{figure*}

A comparison of the diameter for both sequential and simultaneous scenarios and all three phantoms to the ground truth diameter of the CAD sketches is shown in Fig.~\ref{fig:Figure15}. The statistical values of the mean absolute error and standard deviation are presented in Table~\ref{tab:Table3}. The results for the stenosis phantom are divided into the wide and narrow part. For both scenarios the mean absolute error is between \SI{0.07}{\milli\meter} (relative error 2.8\%) and \SI{0.15}{\milli\meter} (relative error 6.0\%) with a standard deviation of \SI{0.01}{\milli\meter} to \SI{0.04}{\milli\meter}. The mean absolute error for the Z-shape and U-shape phantom ranges between \SI{0.06}{\milli\meter} (relative error 2.4\%) and \SI{0.21}{\milli\meter} (relative error 8.3\%) with a standard deviation between \SI{0.07}{\milli\meter} and \SI{0.16}{\milli\meter}. Additionally, we performed a paired t-test to investigate the significance of the differences between sequential and simultaneous imaging with Julia's \textit{HypothesisTests} package. The null hypothesis that the differences between pairs of values come from a distribution with mean $\mu_0$ can be rejected for all phantoms with a significance level of 5\% since the all p-values are below the significance level.

\def\sOffset{-2.75}

\begin{table}[!tb]
\centering
\caption{Mean absolute error (MAE) and standard deviation $\sigma$ of IVOCT diameter to CAD diameter for the sequential and simultaneous method and three phantoms stenosis (wide), stensosis (narrow), Z-shape, and U-shape. P-values are calculated with the paired t-test for differences of the diameters for both methodologies.}
\label{tab:Table3}
\begin{tabular}{ l  l p{1.2cm} p{1.2cm}  c  c }
  \hline \hline
  Phantoms & & Stenosis (wide) & Stenosis (narrow) & Z-shape & U-shape  \\ \hline
  Sequential & MAE [\SI{}{\milli\meter}] & 0.15 & 0.13 & 0.11 & 0.21 \\ 
  Sequential & $\sigma$ [\SI{}{\milli\meter}] & 0.04 & 0.03 & 0.12 & 0.16 \\ 
  Simultaneous & MAE [\SI{}{\milli\meter}] & 0.12 & 0.07 & 0.06 & 0.14 \\  
  Simultaneous & $\sigma$ [\SI{}{\milli\meter}] & 0.04 & 0.01 & 0.07 & 0.08 \\ 
  p-value (paired t-test) & & $10^{-99}$ & $10^{-63}$  & $10^{-99}$ & $10^{-99}$ \\ \hline \hline
\end{tabular}
\end{table}

\subsubsection{In-vivo Feasibility Study}
\begin{figure}
\centering
\includegraphics[width=0.5\linewidth]{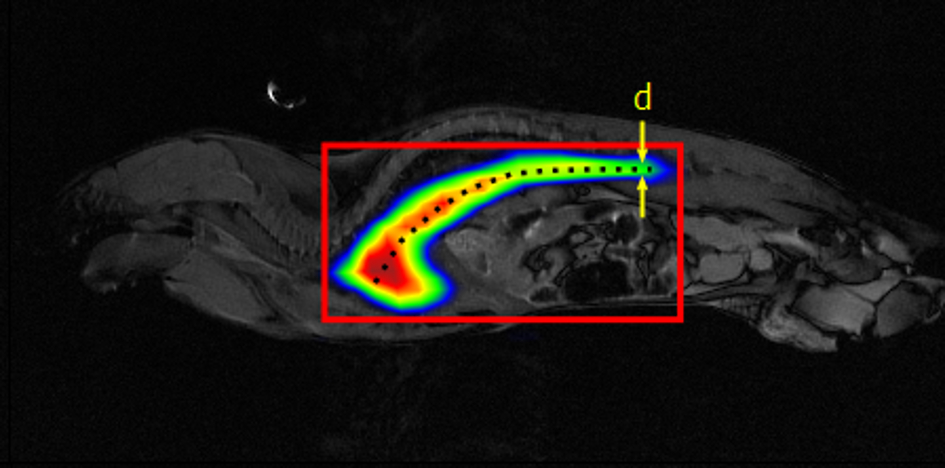}
\caption{Sagittal slice of overlayed MRI and MPI images of a mouse with injected MPI sensitive SPIOs. The MPI FoV with \SI{19}{\milli\meter}~x~\SI{37}{\milli\meter} is highlighted with a red frame. The intensity values of the MPI signal are visualized with different colors (red - high intensity, blue - low intensity). The transition of particles from the vena cava to the mouse heart is visualized. A centerline of the visualized vena cava is approximated and indicated with black dots. At the smallest visible part of the vena cava a diameter $d$ of \SI{0.8}{\milli\meter} is measured (arrows).}
\label{fig:Figure16}
\end{figure}

The MPI and MRI images of the mouse model are overlayed in Figure~\ref{fig:Figure16}. The organs and the skeleton are visible in the gray value MRI image. The injected MPI sensitive SPIOs lead to different MPI signal intensity values and are visualized accordingly, whereas a high concentration of particles per voxel lead to a increased intensity (red). The transition of particles from the vena cava to the mouse heart is shown. The MPI FoV is highlighted with a red bounding box. A centerline of the imaged vessel is indicated with a dashed line. Within the MPI FoV the smallest diameter of the vessel is estimated with $d = \SI{0.8}{\milli\meter}$.

\section{Discussion}
The results show that MPI is a suitable modality to complement IVOCT and to estimate the pullback path of the catheter. Particularly, we demonstrate that imaging can be done sequentially or simultaneously. When the MPI measurement is performed prior to the IVOCT measurement, the vessel could be flushed without SPIOs for the IVOCT measurement. When the measurement is done simultaneously the vessel has to be flushed with SPIOs. We have shown that both methods work similarly well and that the simultaneous measurement with flushed SPIOs have no substantial negative influence on the IVOCT imaging. The SPIO concentration was adjusted to a range where the solution was nearly transparent for IVOCT while still providing a very good MPI signal. With regard to the SPIO studies, simultaneous measurements may be preferable, since it would allow for taking vessel movements into account that could be synchronized between the MPI and IVOCT data.

When comparing the three different volume reconstruction methods, it is clear that the IVOCT image volume based on the MPI trajectory estimate results in a much improved representation of the true vessel shape. Methods A and B rely on the IVOCT data only and show substantial deviations from the ground-truth CAD phantom models, especially for the curved phantoms. As the catheter position was not centered within the phantom most of the time, twists and bends alongside the volume appear for method A. While the center-of-mass approach in method B leads to smoother surfaces, it still does not reflect the actual curvature of the phantoms. Method C combines IVOCT and MPI data and the improved conformity to the CAD model is shown in Fig.~\ref{fig:Figure14}.

The good agreement with respect to the shape is also supported by the quantitative analysis of trajectory and diameter. Table \ref{tab:Table2} shows that MPI can effectively estimate the centerline of the vessel phantoms with a mean absolute error of less than \SI{0.3}{mm}. This is comparable to angiography \cite{yang14,kunio17}. A comparison of the p-values determined with paired t-test show no significance with a significance level of 5\% between the sequential and simultaneous methods in terms of estimating the pullback pathway. 
The errors in the IVOCT estimated vessel diameter are low, ranging from \SI{0.06}{mm} to \SI{0.21}{mm} and are likewise comparable to angiography. The rather precise estimation of the diameter is also illustrated in Fig.~\ref{fig:Figure15}. Note that the errors are computed with respect to the CAD models of the phantom and thus can be considered end-to-end errors including errors from 3D printing, registration, and segmentation. Further, the results of the performed t-test show that the difference between the sequential and simultaneous imaging method are significant for all three phantoms. Since the overall mean absolute errors are lower for the simultaneous method and its results are significantly different from the sequential method, it is indicated that the simultaneous imaging method performs slightly better than the sequential method. Nevertheless, due to the small mean absolute error for sequential and simultaneous both methodologies work within the error order of \SI{0.21}{\milli\meter}-\SI{0.06}{\milli\meter} sufficiently well.

A larger deviation in the estimated diameter was found in the curved parts of the phantoms where the catheter had to bend. These deviations can be explained by the fact that our current approach did not estimate the gradient along the trajectory. Hence, the deviation could be mitigated by placing the B-scans perpendicular to the 3D estimated pathway as illustrated in Fig.~\ref{fig:Figure08} (middle and right). Furthermore, the actual tip motion may not follow the vessels centerline. While this was outside the scope of our feasibility study, it would be possible to label the catheter with a specific MPI marker to track its motion inside a vessel, e.g., by using different magnetic characteristics allowing for simultaneous multi-spectral MPI imaging of marker and vessel \cite{haegele2016multi}.

One advantage of an improved spatial reconstruction of IVOCT is its more realistic representation of distances. Particularly, machine learning approaches working on the raw image data may implicitly use the size of lesions as a feature for classification \cite{gessert2018force}. When the actual shape information is neglected, such features cannot be derived, or worse, irregular patterns in the A-scan alignment may add to overfitting.

In an \textit{in-vivo} study we have illustrated the feasibility of MPI to image vessels in a mouse model. The shape of the vena cava and adjacent heart is visible in MPI. Although the MPI resolution is not high enough to image smaller vessels, e.g. coronary arteries or even the aortic arch, we have shown that an estimation of a vessel centerline for the vena cava is possible \textit{in-vivo}. Note, that we cannot obtain a quantitative analysis of the vessel shape, as there is no ground truth. Moreover, an \textit{in-vivo} study of bimodal IVOCT and MPI imaging is not feasible as the used imaging catheter with diameter of \SI{0.9}{\milli\meter} is too large to be inserted in mouse vessels. 
A practical advantage of using MPI to estimate vessel shape and catheter trajectory is its non-ionizing nature. In contrast to risks of DSA \cite{katzberg2006contrast,mccullough1997acute,mccullough2008contrast}, MPI could continuously monitor the vessel shape and catheter motion for long periods, particularly when long circulating tracers are used \cite{kaul2017vitro}. Another advantage of combining MPI and IVOCT is the high temporal resolution \cite{knopp2017recent, knopp2017magnetic,Wang2015}, allowing to monitor and compensate motion. Clearly, so far our experiments have neglected any vessel motion. However, we have shown that the MPI volume represents the vessel shape and we still have the potential to fully exploit the high temporal resolution of \SI{46}{\hertz} of MPI. In the future, MPI has to prove its potential in a clinical setup with a human study but to this date only preclinical scanners are available. After development of a human sized MPI scanner the presented method of combining MPI and IVOCT has to be validated within a clinical study.

\section{Conclusion}

We have analyzed different MPI tracer concentrations and demonstrated the feasibility to perform simultaneous bimodal IVOCT and MPI imaging. The two modalities have different imaging depth and resolution, i.e., the very high spatial resolution of IVOCT is combined with MPI's ability to show the vessel shape in patient coordinates, which is also demonstrated in an \textit{in-vivo} feasibility study, without using ionizing radiation. We have successfully used MPI based 3D catheter trajectory estimation to improve IVOCT image reconstruction. In conclusion, both modalities complement each other and allow radiation free acquisition of high resolution images of a vessel's shape and morphology.

\appendix

\acknowledgments 
F.G., M.M., M.G., and T.K. thankfully acknowledge the financial support by the German Research Foundation (DFG, grant number KN 1108/2-1) and the Federal Ministry of Education and Research (BMBF, grant number 05M16GKA). A.S. acknowledges partial support by the German Research Foundation (DFG, grant number SCHL 1844/2-1).

\section*{Disclosure of Conflicts of Interest}
The authors have no conflicts to disclose.

\end{spacing}

\bibliographystyle{ama}
\bibliography{ref}

\end{document}